\documentclass[manuscript,nonacm]{acmart}
\AtBeginDocument{%
  \providecommand\BibTeX{{%
    \normalfont B\kern-0.5em{\scshape i\kern-0.25em b}\kern-0.8em\TeX}}}

\usepackage{graphicx}
\usepackage[normalem]{ulem}
\useunder{\uline}{\ul}{}
\usepackage{multirow}
\usepackage{adjustbox}
\usepackage{caption,subcaption,graphicx}

\usepackage{times,tcolorbox}
\usepackage{xcolor}
\usepackage{tcolorbox}

\definecolor{airforceblue}{rgb}{0.36, 0.54, 0.66}
\definecolor{bubblegum}{rgb}{0.99, 0.76, 0.8}
\definecolor{babyblue}{rgb}{0.54, 0.81, 0.94}
\definecolor{bananamania}{rgb}{0.98, 0.91, 0.71}
\definecolor{cobalt}{rgb}{0.0, 0.28, 0.67}

\tcbset{on line, 
        boxsep=1pt, left=0pt,right=0pt,top=0pt,bottom=0pt,
        colframe=white,colback=bubblegum}
        
\newtcbox{\bluebox}{on line, 
        boxsep=1pt, left=0pt,right=0pt,top=0pt,bottom=0pt,
        colframe=white,colback=babyblue}

\newtcbox{\yellowbox}{on line, 
        boxsep=1pt, left=0pt,right=0pt,top=0pt,bottom=0pt,
        colframe=white,colback=bananamania}

\newtcbox{\pinkbox}{on line, 
        boxsep=1pt, left=0pt,right=0pt,top=0pt,bottom=0pt,
        colframe=white,colback=bubblegum}

\begin{document}

\title[Improving Health Professionals' Onboarding with AI and XAI for Trustworthy Human-AI Collaborative Decision Making]{Improving Health Professionals' Onboarding with AI and XAI for Trustworthy Human-AI Collaborative Decision Making}

\author{Min Hun Lee}
\email{mhlee@smu.edu.sg}
\affiliation{%
  \institution{Singapore Management University}
  \city{Singapore}
  \country{Singapore}
}

\author{Silvana Choo Xinyi}
\affiliation{%
  \institution{Singapore General Hospital}
  \city{Singapore}
  \country{Singapore}
}

\author{Shamala D/O Thilarajah}
\affiliation{%
  \institution{Singapore General Hospital}
  \country{Singapore}}

\renewcommand{\shortauthors}{Lee et al.}


\begin{abstract}
With advanced AI/ML, there has been growing research on explainable AI (XAI) and studies on how humans interact with AI and XAI for effective human-AI collaborative decision-making. However, we still have a lack of understanding of how AI systems and XAI should be first presented to users without technical backgrounds. In this paper, we present the findings of semi-structured interviews with health professionals (n=12) and students (n=4) majoring in medicine and health to study how to improve onboarding with AI and XAI. For the interviews, we built upon human-AI interaction guidelines to create onboarding materials of an AI system for stroke rehabilitation assessment and AI explanations and introduce them to the participants. Our findings reveal that beyond presenting traditional performance metrics on AI, participants desired benchmark information, the practical benefits of AI, and interaction trials to better contextualize AI performance, and refine the objectives and performance of AI. Based on these findings, we highlight directions for improving onboarding with AI and XAI and human-AI collaborative decision-making.
\end{abstract}



\maketitle

\section{Introduction}
Artificial intelligence (AI) has been increasingly being explored to provide data-driven insights for improving various decision-making tasks (e.g. health \cite{beede2020human,cai2019human,lee2021human,wang2021brilliant,caruana2015intelligible} and other social services \cite{kuo2023understanding,zavrvsnik2020criminal}). 
Even if recent research has demonstrated that these AI systems can have competent performance that can rival domain experts \cite{cai2019human,topol2019high,lee2019learning,nam2019development,singh2018deep,esteva2017dermatologist}, a fully autonomous approach of AI systems in high-stake contexts (e.g. health) is not desirable due to safety and ethical issues. A growing body of research has been conducted to investigate how humans and AI systems can complement each other's strengths \cite{cai2019human,topol2019high,lee2021human} and integrate these AI systems in practice \cite{beede2020human,nam2019development,singh2018deep}. However, it is still challenging to integrate these systems in practice \cite{sutton2020overview,khairat2018reasons,kohli2018cad,wang2021brilliant} due to several factors, such as lack of user acceptance and trust \cite{sutton2020overview,khairat2018reasons,kohli2018cad} and difficulty with understanding the rationale of an AI output \cite{london2019artificial,rajpurkar2022ai,cai2019human}.

To address these challenges of integrating AI systems in practice, there is growing work that aims to make AI human-centered \cite{cai2019hello,lee2020co,beede2020human}, explainable \cite{arya2019one,wang2019designing,lakkaraju2020explaining,abdul2018trends,preece2018asking}, and trustworthy \cite{floridi2019establishing,liang2022advances,araujo2020ai}. Along this line, recent research works involved stakeholders to understand their practices and needs \cite{lee2020co,yang2019unremarkable,wang2021brilliant} and socio-environmental factors \cite{beede2020human} to design explainable AI techniques to provide new insights on a decision making task and study how clinicians or health professionals can make use of AI outputs \cite{cai2019effects,lee2020co,cai2019human}. However, previous studies \cite{lee2020co,cai2019human,bussone2015role} assume that users without technical backgrounds can be onboarded with an AI system and AI explanations. Some research described the failures of effectively using AI explanations as they might be inadvertently the most understandable for users with technical backgrounds \cite{suresh2021beyond}. Additionally, there has been limited understanding of how AI systems should be introduced to users without technical backgrounds \cite{cai2019hello} and whether they can specify a desirable basic performance of AI to consider using it.

In this work, we focus on the context of physical stroke rehabilitation assessment and explore how AI and AI explanations should be introduced to users without technical backgrounds (e.g. health professionals and students majoring in medicine and health). To this end, we leveraged previous research of guidelines for human-AI interaction \cite{GooglePAIR2019,amershi2019guidelines}, AI model card \cite{mitchell2019model}, onboarding recommendations \cite{cai2019human}, and tutorials of XAI techniques \cite{lakkaraju2020explaining} to create onboarding tutorial materials of an AI-based decision support system and XAI techniques for the context of the study.

Our onboarding tutorial materials include 1) the description of the context (Figure \ref{fig:tutorial_context1} and \ref{fig:tutorial_context2}) and primary usage of AI (Figure \ref{fig:tutorial_apps}), 2) the introduction of AI (i.e. Figure \ref{fig:tutorial_ai}: inputs and outputs of AI \& how it can be developed and operate, Figure \ref{fig:tutorial_data}: dataset, and Figure \ref{fig:tutorial_performance} performance metrics and performance of AI), and 3) the descriptions of the motivation and meaning of an AI explanation (Figure \ref{fig:tutorial_xai}) and three widely used AI explanations (i.e. feature importance, counterfactuals, and prototype/example-based) for the context of the study (Figure \ref{fig:tutorial_feat}, \ref{fig:tutorial_counter}, and \ref{fig:tutorial_example}). 

Using our onboarding tutorial materials, we conducted a semi-structured interview with 12 health professionals and 4 students majoring in medicine and health. Throughout the interview, we learned their practices to build a trustworthy relationship with their colleagues. We also collected participants' feedback about the tutorial materials including their confusions to understand their information needs and areas for improvement. In addition, they were asked to describe a desirable performance AI to consider using it, rank the usefulness of three AI explanations for onboarding and decision support, and share suggestions on how to improve the onboarding and decision support with AI and AI explanations.

Our findings highlight the value of tutorials on AI and AI explanations along with the information needs of users without technical backgrounds (e.g. health professionals and students in medicine and health) on functional, developmental, and evaluation aspects of AI and how to make use of AI explanations. Specifically, participants suggested the context-specific required AI performance and evaluations to determine the usage of AI. Beyond presenting a numerical traditional performance metric, they also recommended communicating the benchmark information and the benefit of AI to contextualize AI capabilities and limitations. As they build a trustworthy relationship with their colleagues over time, they suggested providing iterative trials to refine AI objectives and tune it with feedback for trustworthy interactions with AI. Additionally, our study uncovered challenges of how AI explanations can be designed and used to improve onboarding with AI and support interactive communications with AI: creating a way to measure the level of understanding of AI and AI explanations, aligning goals between users and AI, and specifying the practices to audit AI.

Overall, our study provides insights into how AI and AI explanations can be presented to users without technical background and contributes to design considerations and challenges to improve their onboarding with AI. Our work advances ongoing discussions around onboarding and education of non-technical, domain users with AI \cite{cai2019hello,kuo2023understanding,shen2020designing} for effective human-AI collaborative decision-making in various high-stake domains.

\section{Related Work}
\subsection{Challenges of Deploying AI-based Decision Support Systems for Human-AI Collaboration}
As AI achieves high performance to replicate expert's decision-making \cite{lee2019learning,nam2019development,singh2018deep,esteva2017dermatologist}, such as diagnosing prostate cancer \cite{nam2019development} or assessing the quality of post-stroke rehabilitation exercises \cite{lee2019learning}, AI has been investigated in the form of a decision support system  \cite{cai2019human,lee2021human}. Specifically, ongoing research efforts explore to integrate AI-based decision support systems that provide data-driven insights (e.g. quickly retrieving similar cases from previously diagnosed patients \cite{cai2019human} and identify important input features \cite{lee2021human}) to enhance domain experts' accuracy and efficiency of decision making into practice \cite{cai2019human,lee2021human,beede2020human,wang2021brilliant}. 
However, it remains challenging to integrate these systems into practice \cite{sutton2020overview,khairat2018reasons,kohli2018cad,wang2021brilliant}. One impediment to adopting these systems has been the lack of user acceptance and trust \cite{sutton2020overview,khairat2018reasons,kohli2018cad}. As these systems utilize a complex AI algorithm and often operate as a black box \cite{sutton2020overview,rajpurkar2022ai,cai2019human}, users have difficulty with understanding why the system provides a certain outcome \cite{london2019artificial,rajpurkar2022ai,cai2019human}. If domain experts (e.g. clinicians, health professionals) do not understand the intended use, functionalities, or capability of an AI-based decision support system \cite{maddox2019questions,sutton2020overview}, they may resist and abandon its usage \cite{khairat2018reasons}.

\subsection{Towards Human-Centered, Trustworthy, and Explainable AI}
Researchers have emphasized the importance of making AI human-centered \cite{cai2019hello,lee2020co,beede2020human} , trustworthy \cite{floridi2019establishing,liang2022advances,araujo2020ai,jacovi2021formalizing}, explainable \cite{arya2019one,wang2019designing,lakkaraju2020explaining,abdul2018trends,preece2018asking} to make it more deployable in practical settings. 
In the following subsections, we summarize the prior work on human-centered, trustworthy, explainable AI and describe how we build upon and differentiate with the prior work. 

\subsubsection{Designing Human-Centered AI}
For human-centered designs and evaluation of AI, increasing recent research works \cite{beede2020human,lee2020co,sendak2020human,yang2019unremarkable} highlight the importance of involving stakeholders to understand their challenges and needs \cite{lee2020co,yang2019unremarkable,wang2021brilliant} and socio-environmental factors \cite{beede2020human}. For instance, Wang et al. \cite{wang2021brilliant} conducted interviews with clinicians in China and conducted observations to examine how AI-based decision support systems are used and discussed the issue of misalignment with local context and workflow and usability barriers. Lee et al. \cite{lee2020co} interviewed and conducted a focus group session with therapists to understand their challenges and needs during rehabilitation assessment to design a human-centered decision support system. Beede et al. interviewed and observed the eye-screening workflows of clinics in Thailand, characterized the user expectations and post-deployment experiences of the AI-assisted screening process, and discussed the necessity of evaluating a system in socio-technical contexts \cite{beede2020human}. 

In this work, we focus on the context of an AI-based decision support system for assessing physical stroke rehabilitation assessment. Building upon a growing body of research that highlights the value of human-centered approaches to invite feedback on designs of AI systems from the target users \cite{lee2020co,yang2019unremarkable,wang2021brilliant,beede2020human}, we engaged with stakeholders without technical backgrounds (e.g. health professionals) to explore how to improve their onboarding with an AI-based decision support system. As stakeholders without technical backgrounds may not provide sufficiently detailed suggestions on a narrow design aspect (e.g. the overall problem formulation) \cite{kuo2023understanding},  
our interdisciplinary team of a technical researcher and domain experts in stroke rehabilitation has worked together to create onboarding materials of AI and AI explanations and conducted semi-structured interviews with health professionals to collect their critiques and suggestions on how to improve onboarding with AI and AI explanations. 

\subsubsection{Efforts on Framework for Trustworthy AI}

Trust is considered as a critical component of the successful deployment of AI and increasing research discusses about creating trustworthy AI \cite{wing2021trustworthy,vashney2022trustworthy,floridi2019establishing,toreini2020relationship}. Although there has been little common understanding of what constitutes trust or trustworthy AI \cite{gille2020we}, researchers have discussed several definitions and frameworks of trustworthy AI. For instance, Vashney \cite{vashney2022trustworthy} builds upon the definition of trust and describes four attributes of trustworthy artificial intelligence: 1) technical competence that refers to the basic performance and accuracy of an AI model, 2) reliability and fairness that indicates maintaining good and correct performance across varying operating conditions,  3) understandability that describes whether users can comprehend the pipeline and lifecycles of an AI model, and 4) personal attachment/benevolence, which refers whether the purpose of an AI model can be aligned with a society’s wants. In addition, Toreini et al. \cite{toreini2020relationship} based on the widely accepted principles of trust, ABI (Ability, Benevolence, Integrity)  \cite{mayer1995integrative} and described a framework of trustworthy AI that includes a temporal dimension from initial trust to continuous trust \cite{li2008we,toreini2020relationship} and four technologies: fairness, explainability, auditability, and safety.

Although there are increasing efforts to make frameworks for trustworthy AI, there are still remaining questions  on how these frameworks can be applied to create a new application of trustworthy AI (e.g. how we can effectively build initial trust with AI? what would be desirable basic performance of trustworthy AI and role of explainable AI techniques?). In addition, most prior work of designing human-centered, AI-based clinical decision support systems
assumed that clinicians or health professionals onboard with AI and then studied how they can interact with these AI-based systems \cite{beede2020human,lee2020co,sendak2020human,yang2019unremarkable,wang2021brilliant}. The problem of how users without technical backgrounds can be onboarded with AI \cite{cai2019hello} (i.e. understandability aspects of trustworthy AI \cite{vashney2022trustworthy}) is underexplored. 

Building upon previous research of trustworthy AI \cite{wing2021trustworthy,vashney2022trustworthy,floridi2019establishing,toreini2020relationship} and guidelines of human-AI interaction \cite{amershi2019guidelines,GooglePAIR2019,cai2019hello,cai2021onboarding} including AI model cards \cite{mitchell2019model}, we explored how onboarding tutorial materials of AI and XAI can be created and presented to users without technical backgrounds for trustworthy AI.
Among various aspects and components of trustworthy AI, our work focuses on exploring how to build initial trust \cite{toreini2020relationship} and effectively onboard with AI while understanding the information needs of health profesionals on AI and XAI and exploring the possibility of defining the user's notion of basic performance of an AI model to start using it.

\subsubsection{Technically Oriented AI Explanations}

To address the user's difficulty with understanding the rationales of an AI output/recommendation, researchers have explored techniques to make AI interpretable and explainable \cite{preece2018asking,arya2019one,wang2019designing,lakkaraju2020explaining}. These explainable AI techniques can be broadly categorized into 1) inherently interpretable models (e.g. rule-based models or linear regressions) whose internal mechanisms are directly interpreted and 2) post-hoc explainable AI (XAI) techniques that provide explanations of a complex algorithm (e.g. a deep learning model) \cite{lakkaraju2020explaining}. Various post-hoc XAI techniques can be further classified into explaining the model's overall or instance-specific behavior \cite{lakkaraju2020explaining}. Among various post-hoc XAI techniques, this work focuses on three widely used local XAI techniques: feature importance, counterfactual, and prototype/example-based explanations. A feature importance explanation describes how much input features contribute to a model output \cite{gilpin2018explaining,ribeiro2016should,mundhenk2019efficient}. A counterfactual explanation describes how input features should be changed to update an AI output \cite{verma2020counterfactual,guidotti2019factual,mothilal2020explaining}. A prototype/example-based explanation aims to identify samples that are the most relevant and influential to an AI output \cite{gilpin2018explaining,cai2019effects}.

Explainable AI techniques that generate rationales of an AI output aim to serve a variety of users: technical AI/ML developers, who monitor and debug an AI model, or a non-technical, domain users, such as clinicians or health professionals, who review AI explanations as relevant evidence and outcomes on a decision-making task. However, prior research has shown that these AI explanations are not useful for people without technical background (e.g. clinicians or health professionals in clinical practice) \cite{bussone2015role,poursabzi2021manipulating}. These failures of effectively using AI explanations might have occurred because these AI explanations are not designed for specific end-users or tasks \cite{suresh2021beyond}. These AI explanation methods might be inadvertently the most understandable to people with technical backgrounds who build and debug an AI model. As the end-users might have different needs, goals, and tasks when interpreting and reacting to AI model outputs and explanations, it is critical to engage with the end-user and make AI explanations user-centered. 

In this work, we utilized three widely used XAI techniques and explored how these techniques can be used to improve users' onboarding with AI for their AI-assisted decision-making. 

The most relevant research to our work is research by Cai et al. \cite{cai2019hello} that describes pathologists' information needs on an AI model (i.e. known strengths and limitations and its design objective). Although the previous work \cite{cai2019hello} provides several suggestions on onboarding with AI, it remains unclear how we can introduce users without technical backgrounds to the functionality, strengths and limitations, and design objectives of AI and AI explanations. Building upon this previous research \cite{cai2019hello}, our research further investigates the usefulness of an AI model card \cite{mitchell2019model} to communicate the competence of an AI model \cite{GooglePAIR2019} for user's onboarding with AI. Specifically, we studied whether users without technical backgrounds (i.e. health professionals) can leverage a traditional performance metric from an AI model card to understand the strengths and limitations of an AI model and determine whether an AI model can be used in practice. In addition, we conducted a deeper examination of aspects to faciliate users' onboarding with AI and AI explanations and how three widely used AI explanations can be used for onboarding and decision-making with AI. Our work further discusses considerations to improve onboarding with AI and AI explanations and human-AI collaborative decision-making.

\section{Study Design}
This work aims to understand how an AI-based decision support system can be introduced to medical practitioners for its trustworthy usage. Specifically, we focused on studying (1) how well medical practitioners can understand the onboarding tutorial materials of an AI decision support system and (2) whether they can leverage a traditional evaluation metric which is commonly used by AI/ML researchers to indicate how well an AI/ML model can classify/predict ground truth scores (e.g. F1-scores) to determine the usage of the system, (3) the usefulness of three AI explanations for onboarding and decision support, and (4) informing the design of onboarding tutorial materials and considerations for trustworthy usage of an AI-based decision support system. 

To address these research questions, we leveraged existing guidelines for human-AI interaction \cite{amershi2019guidelines,GooglePAIR2019} and onboarding with AI \cite{cai2019hello}, an AI model card \cite{mitchell2019model}, tutorials of AI explanations \cite{lakkaraju2020explaining} to create the onboarding tutorial materials of an AI-based decision support system for physical stroke rehabilitation assessment (Figure \ref{fig:tutorial_contexts}, \ref{fig:tutorials_ai}, \ref{fig:tutorials_xai_all}). In addition, we had iterative online synchronous and asynchronous discussions with domain experts in stroke rehabilitation to inform a set of semi-structured interview questions and refine onboarding tutorial materials of an AI-based decision support system for physical stroke rehabilitation assessment. During the online synchronous discussions, the leading researcher with a background of human-AI interaction and machine learning presented the draft of interview questions and onboarding materials and collected feedback on areas to be improved and revised questions, onboarding materials, and scripts for the follow-up discussions. After refining and finalizing the interview questions and onboarding materials, we conducted a pilot interview session with a student who majors in law and does not have technical backgrounds to check the length of a session and whether onboarding materials are understandable for people without technical backgrounds. Then, we conducted a semi-structured interview with healthcare professionals (i.e. therapists and a medical social worker) and students majoring in medicine and healthcare (e.g. nursing, therapy). This study including onboarding tutorial materials, protocol, and recruitment methods was approved by the Institutional Review Board. 

\subsection{Onboarding Tutorial Materials}

Our onboarding tutorial materials are composed of three parts: introducing 1) the context and AI applications for physical stroke rehabilitation, 2) the development and evaluation of an AI model (e.g. how it is trained and operates on new data, dataset, evaluation metrics, and performance), 3) AI explanations (e.g. the motivation of AI explanations, feature importance, counterfactual, and example/prototype-based explanations).

First, building upon guidelines \cite{GooglePAIR2019,mitchell2019model}, we described the context and challenges that an AI-based system aims to address (Figure \ref{fig:tutorial_context1}), the primary applications for the users of this study (i.e. therapists and post-stroke survivors) (Figure \ref{fig:tutorial_apps}), and envisioned use cases of quantitative stroke rehabilitation assessment (i.e. assessing the range of motion, smoothness, and the presence of compensatory motions) (Figure \ref{fig:tutorial_context2}).
The descriptions of tutorial materials (Figure \ref{fig:tutorial_contexts}) to introduce the contexts of the study can be found below:

\begin{figure*}[htp]
\centering 
\begin{subfigure}[t]{0.7\textwidth}
\centering
  \includegraphics[width=1.0\columnwidth]{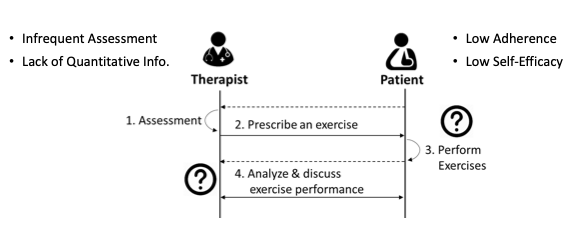}
  \caption{}
  \label{fig:tutorial_context1}
\end{subfigure}\hspace{5mm}
\begin{subfigure}[t]{0.8\textwidth}
  \centering
  \includegraphics[width=1.0\columnwidth]{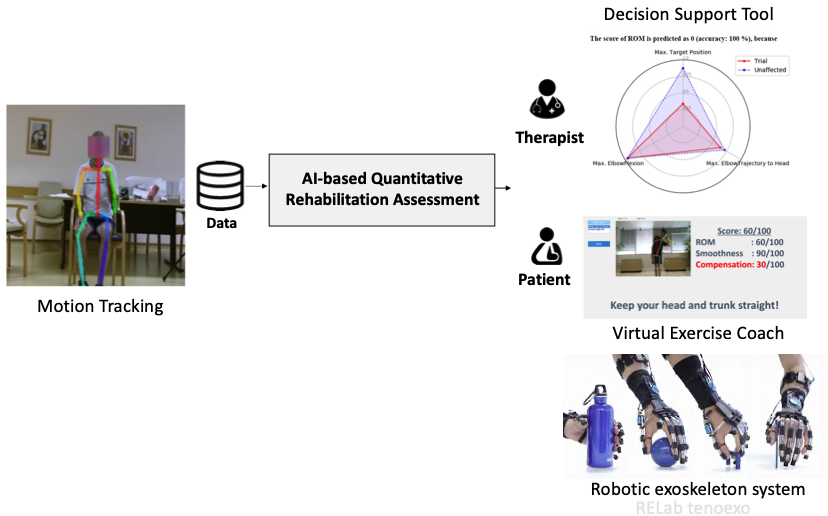}
  \caption{}
  \label{fig:tutorial_apps}
\end{subfigure}
\begin{subfigure}[t]{0.8\textwidth}
  \centering
  \includegraphics[width=1.0\columnwidth]{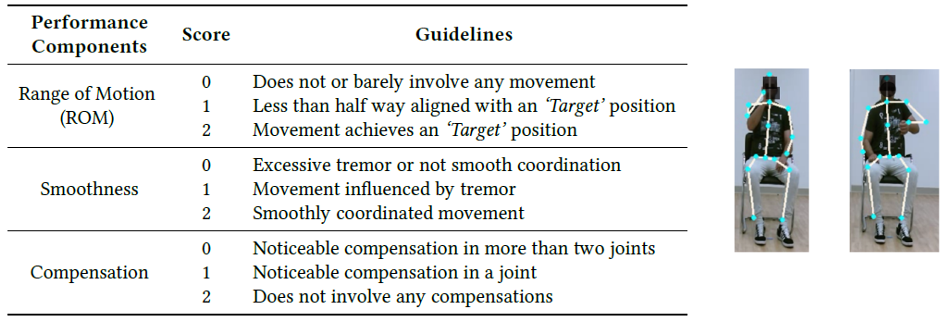}
  \caption{}
  \label{fig:tutorial_context2}
\end{subfigure}
\caption{Onboarding Tutorial Materials of an AI that introduce (a) an (c) the context of physical stroke rehabilitation assessment and (b) AI applications of physical stroke rehabilitation assessment and therapy.}\label{fig:tutorial_contexts}
\end{figure*}

\begin{quote}
\small{
    Figure \ref{fig:tutorial_context1}:
    \textit{``When stroke occurs, post-stroke survivors will have paralyzed and limited functional abilities. They typically involve therapy sessions to regain their functional and cognitive abilities. During therapy sessions, therapists assess the functional \& cognitive status of a patient and prescribe a set of exercises to practice.  In this work, we focus on the functional assessment of post-stroke survivors. During rehabilitation therapy, therapists often prescribe a set of exercises to a patient due to their limited availability. A therapist and a patient have regular follow-up meetings to discuss the patient’s status and progress and adjust the rehabilitation program accordingly. During the follow-up meeting, there is limited quantitative information on the patient’s status for therapists to make informed decision making.''}   
}
\end{quote}

\begin{quote}
\small{
    Figure \ref{fig:tutorial_apps}:
    \textit{``To address this challenge, there have been increasing explorations on an AI-based system for rehabilitation. This system typically utilizes a vision-based or wearable sensor to estimate body joint positions and extract various kinematic features to quantify the patient’s quality of motion.  This quantitative assessment can be provided to therapists as a decision support system for improving their rehabilitation assessment or a virtual or exoskeleton system to improve patients’ engagement in rehabilitation.''}
}
\end{quote}

\begin{quote}
\small{
    Figure \ref{fig:tutorial_context2}:
    \textit{``For the rehabilitation assessment, therapists assess patient’s quality of motion in the following three aspects: Range of Motion, Smoothness, Compensation. ROM indicates whether a patients can achieve a specific target motion. Smoothness indicates whether a patients can coordinate their motion smoothly. Compensation checks whether a patient involves any unnecessary joint motion; here this patient compensates with his shoulder and trunk to move his arm that is affected by stroke.''}
}
\end{quote}

In the second part of the onboarding tutorial materials, as suggested by the guidelines \cite{GooglePAIR2019,mitchell2019model}, we explained the inputs and outputs of an AI-based system and how a typical AI-based system for rehabilitation assessment can be developed and operated (Figure \ref{fig:tutorial_ai}) and described the dataset \cite{lee2019learning} (i.e. how it is collected and labeled) 
(Figure \ref{fig:tutorial_data}). 
In addition, we elaborated on performance metrics and the performance of an AI model \cite{GooglePAIR2019,mitchell2019model} (Figure \ref{fig:tutorial_performance}). For reporting the AI performance, we utilized the dataset \cite{lee2019learning} and followed the previous research on quantitative stroke rehabilitation assessment \cite{lee2019learning} to implement a feed-forward neural network model from using Pytorch libraries \cite{paszke2019pytorch}. The implementation details of the AI model can be found in the Appendix. \ref{sect:appendix_aimodel}. We then reported how well an AI model can assess three common performance components of rehabilitation assessment (i.e. range of motion, smoothness, and compensation). The descriptions of tutorial materials (Figure \ref{fig:tutorial_ai}) to introduce the development, operation, and evaluation of AI can be found below:

\begin{figure*}[htp]
\centering 
\begin{subfigure}[t]{0.9\textwidth}
\centering
  \includegraphics[width=1.0\columnwidth]{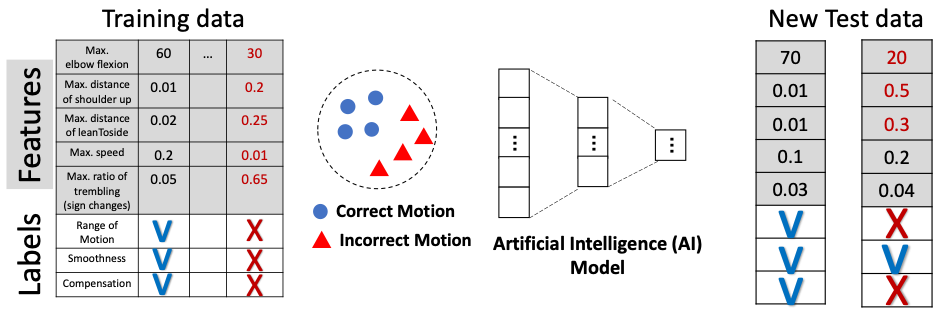}
  \caption{}
  \label{fig:tutorial_ai}
\end{subfigure}\hspace{5mm}
\begin{subfigure}[t]{0.9\textwidth}
  \centering
  \includegraphics[width=1.0\columnwidth]{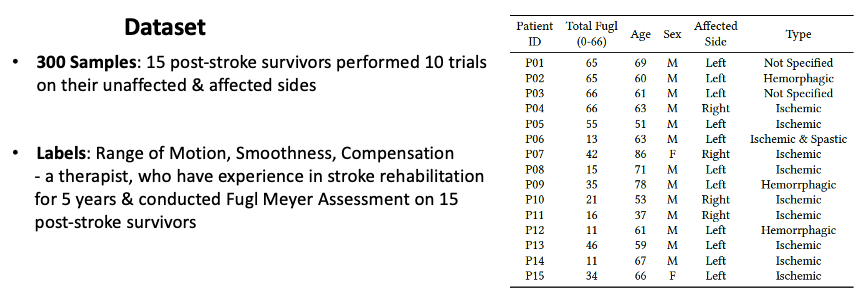}
  \caption{}
  \label{fig:tutorial_data}
\end{subfigure}
\begin{subfigure}[t]{0.9\textwidth}
\centering
  \includegraphics[width=1.0\columnwidth]{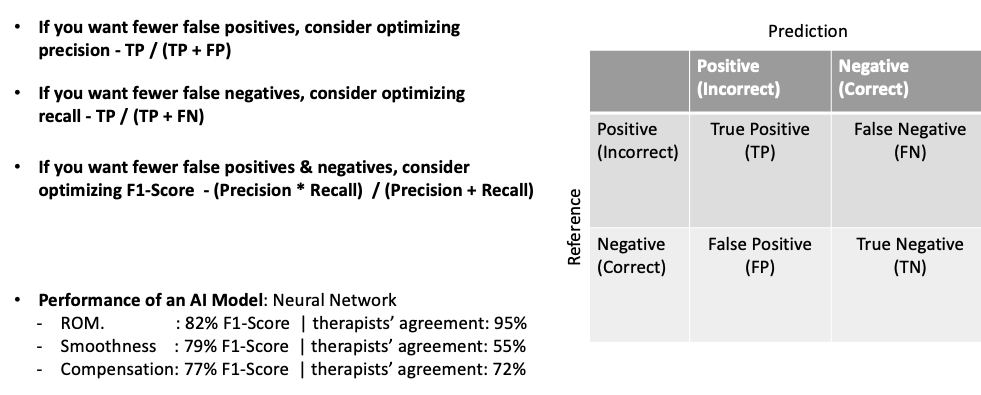}
  \caption{}
  \label{fig:tutorial_performance}
\end{subfigure}\hspace{5mm}
\caption{Onboarding Tutorial Materials of an AI: (a) a diagram of how AI is developed and operated, (b) dataset, and (c) evaluation metrics and performance}\label{fig:tutorials_ai}
\end{figure*}

\begin{quote}
\small{
    Figure \ref{fig:tutorial_ai}:
    \textit{``Here, we describe the pipeline of developing an AI model in more detail. When the system estimates body joints, it extracts kinematic features, such as elbow flexion, how much each joint moves in a certain direction, and computes the overall statistics during an exercise, such as the maximum value of elbow flexion and the corresponding labels of exercises, whether an exercise has the full range of motion or not; smooth or not; involves any compensations or not; \\
    For training an AI model, we collect samples of patients’ exercises, extract features, and collect labels. Here, we have only 5 kinematic features, but in a real case of the development, we have a lot more features and samples. Given these paired features and labels, an AI model learns a function that maps features and corresponding labels as closely as possible. \\
    Given new test data and extracted features from a patient, the AI model generates an outcome whether an exercise has full ROM or not, smooth, or involves any compensations or not. Here, the elbow flexion angle is similar to the normal case. And the distance of shoulder up is also small and close to the normal and along with a similar ratio of trembling. Thus, we have AI outputs of normal ROM, smooth, and no compensation.''}}
\end{quote}

\begin{quote}
\small{
    Figure \ref{fig:tutorial_data}:
    \textit{``For an explorative study, we collected 300 samples of exercises, in which 15 post-stroke survivors performed 10 trials on their unaffected and affected sides. We collected labels of these exercises from a therapist, who has 5 years of practicing stroke rehabilitation and conducted a fugl meyer assessment on 15 post-stroke survivors. Using this dataset, we developed an AI model to replicate therapist’s assessment on ROM, Smoothness, Compensation of patient’s exercises.''}
}
\end{quote}

\begin{quote}
\small{
    Figure \ref{fig:tutorial_performance}:
    \textit{``Any AI model that we build is guided by a reward function, which the AI model uses to determine `right' or `wrong' outcomes. We should consider how we specify this reward function that the system will optimize for \cite{GooglePAIR2019}.\\
    When an AI generates outcomes of whether an exercise is correctly conducted or not, there are four possible outcomes \cite{GooglePAIR2019}:
    True Positive: indicates AI outputs an `Incorrect' motion given an `Incorrect' motion;
    True Negative: indicates AI outputs "correct" motion given a `correct' motion;
    False Positive: indicates AI outputs an `Incorrect' motion given a `correct' motion;
    False Negative: indicates AI outputs `correct' motion given an `Incorrect' motion\\
    If you want fewer false positives, you can consider optimizing precision;
    If you want fewer false negatives, you can consider optimizing recall;
    If you want fewer false positives \& negatives, you can consider optimizing the F1-score\\
    Given our dataset, we optimize an AI model to have fewer false positives \& negatives. 
An AI model achieves a 82\% F1-score on ROM; 79\% on Smoothness; 77\% on compensation; \\
To understand the competence of an AI model to replicate a therapist's assessment, we computed how well a secondary therapist agrees with the therapist, who generated annotations. Overall, our AI model can achieve comparable performance with a secondary therapist.''}}

\end{quote}

In the third part of the tutorial materials, we first described the motivation and meaning of an AI explanation \cite{lakkaraju2020explaining} using an image classification task \cite{ribeiro2016should}. In addition, we explained three commonly used local AI explanations (i.e. feature importance, counterfactuals, and prototypes/example-based) \cite{doshi2017towards,lakkaraju2020explaining} for the context of the study. The descriptions of tutorial materials (Figure \ref{fig:tutorials_xai_all}) to introduce the AI explanations can be found below:

\begin{figure*}[htp]
\centering 
\begin{subfigure}[t]{0.65\textwidth}
  \centering
  \includegraphics[width=1.0\columnwidth]{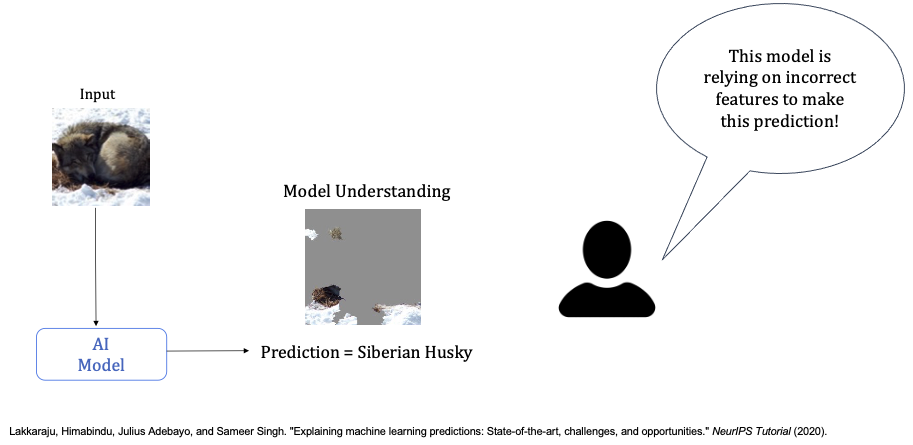}
  \caption{}
  \label{fig:tutorial_xai}
\end{subfigure}
\begin{subfigure}[t]{0.7\textwidth}
\centering
  \includegraphics[width=1.0\columnwidth]{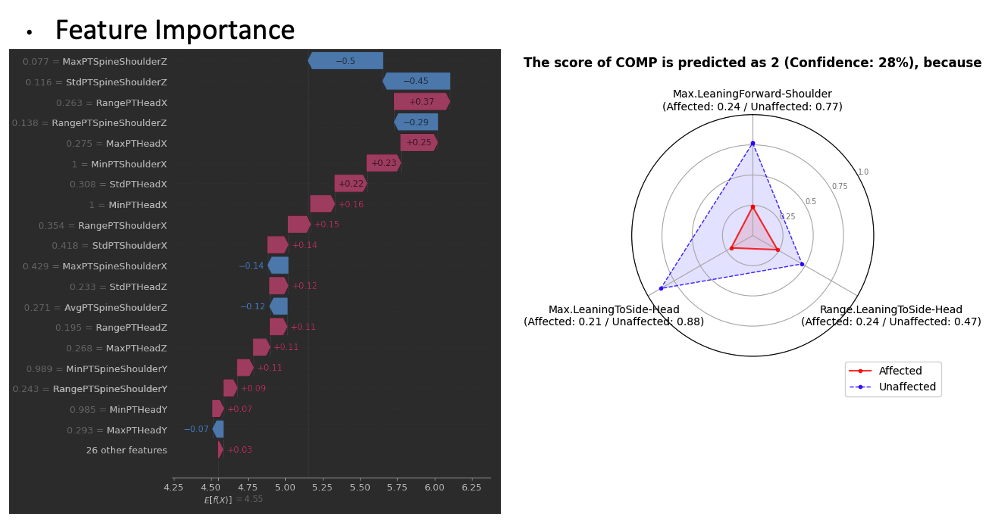}
  \caption{}
  \label{fig:tutorial_feat}
\end{subfigure}\hspace{5mm}
\begin{subfigure}[t]{0.7\textwidth}
  \centering
  \includegraphics[width=1.0\columnwidth]{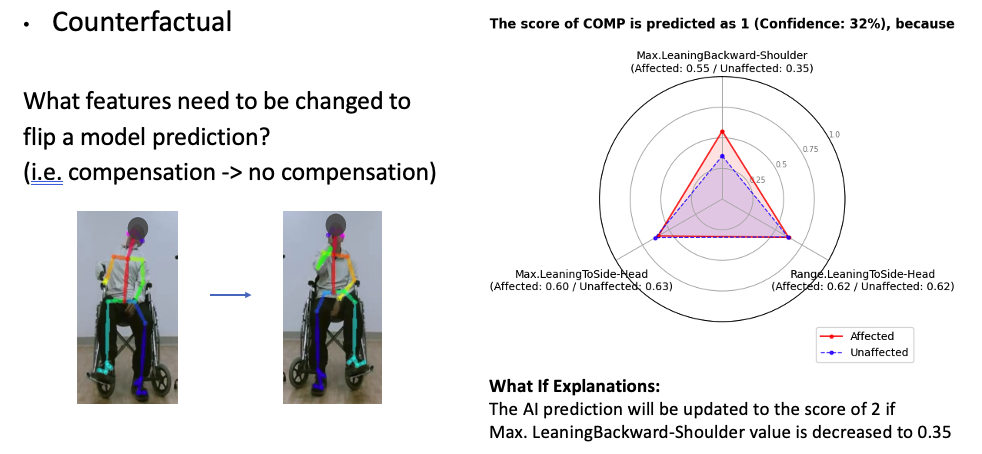}
  \caption{}
  \label{fig:tutorial_counter}
\end{subfigure}
\begin{subfigure}[t]{0.7\textwidth}
  \centering
  \includegraphics[width=1.0\columnwidth]{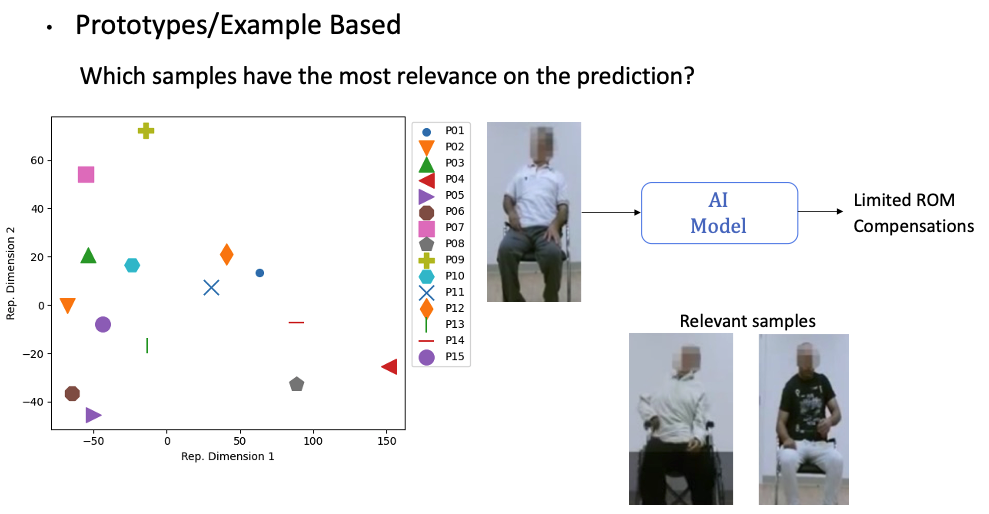}
  \caption{}
  \label{fig:tutorial_example}
\end{subfigure}
\caption{Onboarding Tutorial Materials of an AI explanations: (a) motivation of an AI explanation, (b) a feature importance explanation, (c) a counterfactual explanation, and (d) prototype/example-based explanations.}\label{fig:tutorials_xai_all}
\end{figure*}

\begin{quote}
\small{
    Figure \ref{fig:tutorial_xai}:
    \textit{``There has been increasing research on AI explanations, which aims to improve users’ understanding of how the AI-based system works and determine when to trust an AI output\\
    Here, we have an AI model that classifies the type of animals from an image; Given an input image, an AI model classifies it as Siberian Husky. Here I give you one example of an AI explanation; An AI explanation describes which parts of an image an AI model relied on for its output. We can find that an AI model focuses on snow parts and identify the limitation of an AI model. Thus, we need to be careful of using this AI model.''}}
    
\end{quote}

\begin{quote}
\small{
    Figure \ref{fig:tutorial_feat}:
    \textit{``Feature importance describes the overall importance of different features on AI model outcomes; Once we identify important features, we can pick the top most important features 
and show the comparison of these feature values on patients’ unaffected and affected sides to check the threshold/boundary between unaffected/affected side makes sense or not and determine whether to trust AI or not.''}
}
\end{quote}

\begin{quote}
\small{
    Figure \ref{fig:tutorial_counter}:
    \textit{``Counterfactuals describe how input features need to be changed to generate an opposite outcome. By reviewing which inputs lead to different outcome, we can understand how an AI operates on specific inputs and determine whether we can trust AI or not. For instance, given an AI output of detecting compensation (1), the counterfactual explanations describe that an AI model will generate the output of no compensation if the feature value of max. leaningbackward-shouldervalue is decreased to 0.35.''}
}
\end{quote}

\begin{quote}
\small{
    Figure \ref{fig:tutorial_example}:
    \textit{``Prototype/Example-based Explanation shows which sample data points are the most similar to input data. By reviewing relevant samples and the outputs of AI models on these samples, we can identify whether an AI model has the right outputs or not. Also, we can summarize the input feature into two key dimensions and use these representations to identify which samples are closely aligned and similar with each other. Specifically, we can project the features representation of each patient in the visualization. For instance, by reviewing this Figure, we can understand how each patient has been represented and which patients are represented/considered similar by an AI model and check whether an AI model utilizes correct feature representations or not.''}
}
\end{quote}

For a feature importance explanation, we utilized the SHAP library that support consistency and local accuracy \cite{lundberg2017unified} to compute importance scores (i.e. SHAP values) of each feature (Figure \ref{fig:tutorial_feat}). In addition, we utilized the top three most important features to show the difference between patient's unaffected and affected sides by stroke using a radar chart \cite{lee2021human,amershi2019guidelines}, following the practices of therapists to compare patient's unaffected and affected side \cite{lee2020co}. 

A counterfactual explanation indicates what changes in feature values will lead to updating an AI output in a certain way \cite{mothilal2020explaining,lakkaraju2020explaining,lee2023understanding}. For a counterfactual explanation, we utilized the DiCE library \cite{mothilal2020explaining} to apply a genetic algorithm \cite{olvera2010review} to find counterfactuals close to the query point. We specified the features to be changed in the DiCE library using the identified salient features by the SHAP library and their desired range using patients’
held-out normal data to avoid generating varying and unfeasible explanations. After identifying counterfactual explanations, we generated textual descriptions of the changes in feature values and AI outputs (Figure \ref{fig:tutorial_counter}). For example, Figure \ref{fig:tutorial_counter} describes that the value of \textit{`Max.LeaningBackward-Shoulder'} should be decreased to 0.35 to update the AI output from 1 (i.e. noticeable compensation) to 2 (i.e. no compensation).

A prototype/example-based explanation describes representative or similar samples of a current instance along with AI outputs and ground truths on those samples \cite{cai2019effects,lakkaraju2020explaining,cai2019human} (Figure \ref{fig:tutorial_example}). We utilized the kinematic features of a patient's motion \cite{lee2019learning} and computed the cosine similarity score to identify similar samples, which assist a user in understanding and validating an AI model output. In addition, we utilized a Principal Component Analysis (PCA) \cite{wold1987principal} to reduce the dimension of a feature representation. PCA was utilized because it does not require hyperparameter tuning and is deterministic unlike another widely used dimensional reduction technique, t-Distributed Stochastic Neighbor Embedding (t-SNE) \cite{van2008visualizing}. We then visualized this reduced feature embedding space \cite{boggust2022embedding} for a user to check whether the feature representation of an AI model is valid or not (Figure \ref{fig:tutorial_example}). 

Although we utilized a particular technique/library to identify important features, select counterfactual explanations, and reduce the dimension of a feature representation, this work does not intend to communicate the usage of these specific techniques to create tutorial materials without any considerations. Alternative techniques/libraries can be explored for different applications.

\subsection{Recruitment and Demographics}
We recruited sixteen participants for our study (Table \ref{tab:demographics}). The detailed demographics can be found in Appendix. Table \ref{tab:demographics_detailed}. 

\begin{table}[h]
\centering
\caption{Demographics of Participants: Therapists who have experience in stroke rehabilitation (\bluebox{P1 - P10}) and other health professionals (\yellowbox{P11 - P12}) and students majoring in medicine or health (e.g. therapy, nursing) (\pinkbox{P13 - P16}).}
\label{tab:demographics}
\resizebox{\textwidth}{!}{%
\begin{tabular}{lllll|lllll} \toprule
\textbf{PID} &
  \textbf{Occuptation} &
  \textbf{\# of yrs} &
  \begin{tabular}[c]{@{}l@{}}\textbf{Q. Tech} \\ \textbf{Experience}\end{tabular} &
  \begin{tabular}[c]{@{}l@{}}\textbf{Q. ML} \\ \textbf{Outputs}\end{tabular} &
  \textbf{PID} &
  \textbf{Occuptation} &
  \textbf{\# of yrs} &
  \begin{tabular}[c]{@{}l@{}}\textbf{Q. Tech} \\ \textbf{Experience}\end{tabular} &
  \begin{tabular}[c]{@{}l@{}}\textbf{Q. ML} \\ \textbf{Outputs}\end{tabular}\\ \midrule
\bluebox{P1} &
  PhysioTherapist  &
  7 &
  \begin{tabular}[c]{@{}l@{}}
  4.8 out of 7\end{tabular} &
  3 out of 3 &
  \yellowbox{P11} &
  Speech Therapist &
  5 &
  \begin{tabular}[c]{@{}l@{}}
  4.8 out of 7\end{tabular} &
  2 out of 3 \\
\bluebox{P2} &
  PhysioTherapist  &
  2 &
  \begin{tabular}[c]{@{}l@{}}
  5.2 out of 7\end{tabular} &
  1 out of 3 &
  \yellowbox{P12} &
  Medical Social Worker &
  5 &
  \begin{tabular}[c]{@{}l@{}}
  4.0 out of 7\end{tabular} &
  1 out of 3 \\
\bluebox{P3} &
  PhysioTherapist  &
  8 &
  \begin{tabular}[c]{@{}l@{}}
  4.4 out of 7\end{tabular} &
  2 out of 3 &
  \pinkbox{P13} &
  Student in Occupational Therapy &
  n/a &
  \begin{tabular}[c]{@{}l@{}}
  3.8 out of 7\end{tabular} &
  3 out of 3 \\
\bluebox{P4} &
  PhysioTherapist  &
  11 &
  \begin{tabular}[c]{@{}l@{}}
  5.8 out of 7\end{tabular} &
  2 out of 3 &
  \pinkbox{P14} &
  Student in Speech Therapy &
  n/a &
  \begin{tabular}[c]{@{}l@{}}
  3.8 out of 7\end{tabular} &
  2 out of 3 \\
\bluebox{P5} &
  PhysioTherapist &
  9 &
  \begin{tabular}[c]{@{}l@{}}
  5.4 out of 7\end{tabular} &
  2 out of 3 &
  \pinkbox{P15} &
  Student in Medicine &
  n/a &
  \begin{tabular}[c]{@{}l@{}}
  3.8 out of 7\end{tabular} &
  2 out of 3 \\
\bluebox{P6} &
  PhysioTherapist  &
  30 &
  \begin{tabular}[c]{@{}l@{}}
  5.8 out of 7\end{tabular} &
  2 out of 3 &
  \pinkbox{P16} &
  Student in Nursing &
  n/a &
  \begin{tabular}[c]{@{}l@{}}
  4.0 out of 7\end{tabular} &
  0 out of 3 \\
\bluebox{P7} &
  Occupational Therapist  &
  14 &
  \begin{tabular}[c]{@{}l@{}}
  5.4 out of 7\end{tabular} &
  2 out of 3 &
   &
   &
   &
   &
   \\
\bluebox{P8} &
  Occupational Therapist  &
  11 &
  \begin{tabular}[c]{@{}l@{}}
  6.2 out of 7\end{tabular} &
  0 out of 3 &
   &
   &
   &
   &
   \\
\bluebox{P9} &
  Occupational Therapist  &
  6 &
  \begin{tabular}[c]{@{}l@{}}
  4.4 out of 7\end{tabular} &
  2 out of 3 &
   &
   &
   &
   &
   \\
\bluebox{P10} &
  Occupational Therapist &
  5 &
  \begin{tabular}[c]{@{}l@{}}
  3.2 out of 7\end{tabular} &
  3 out of 3 &
   &
   &
   &
   &
   \\ \bottomrule
\end{tabular}%
}
\end{table}

Our participants were mostly therapists who had experience in stroke rehabilitation (\bluebox{P1 - P10}). Among ten therapists in stroke rehabilitation, six of them are physiotherapists, who promote and maintain patient's physical impairments from bio-mechanical perspectives and four of them are occupational therapists, who assist patients to better engage in their daily activities. Therapists work in various settings: four participants are from outpatient clinics, three participants are from inpatient rehabilitation, two participants are from home care, and two participants are from skilled nursing facility (Appendix. Table \ref{tab:demographics_detailed}). As some outpatient clinics have an interdisciplinary team to support and manage a patient (e.g. physio/occupational therapists, speech therapists, nurses, doctors), we also included health professionals (\yellowbox{P11 and P12}) (e.g. speech therapist and a medical social worker) and students majoring in medicine and health (e.g. therapy and nursing) (\pinkbox{P13 - P16}). The student in occupational therapy (P13) and the student in speech therapy (P14) had an experience of working as an occupational/speech therapy assistant for stroke rehabilitation. Participants were recruited through advertisements sent to the hospital staff, the mailing lists, and the contacts of the research team.

We asked the participants to respond to a set of technical experience questions on recent technologies , which were based on questions designed by the Center for Research and Education on Aging and Technology Enhancement (CREATE) \cite{czaja2006factors}. Specifically, they were asked to rate their experiences with recent technologies (i.e. computer/laptop, activity tracker, virtual voice assistant, unmanned convenient store, and autonomous vehicle) on a 7-point scale (1 = strongly disagree, 2 = disagree, 3 = somewhat disagree, 4 = neutral, 5 = somewhat agree, 6 = agree, and 7 = strongly agree). A high score on technology experience (e.g. 7) indicates that a participant self-reported to be highly experienced with a recent technology. Overall, participants expressed a diverse level of experience with recent technologies, in which they had an average score of 4.31 out of 7.0. In addition, the column of `Q.ML Outputs' in Table \ref{tab:demographics} describes how many times a participant correctly guess an AI output after introducing how an AI model is developed and operated (Figure \ref{fig:tutorial_ai}). In Section \ref{sect:results_info_ai}, we described overall results of participants and how well the scores of guessing AI outputs are correlated with the scores of technology experiences.

\subsection{Protocol}
We conducted semi-structured interviews with 16 participants: 12 healthcare professionals (11 therapists and 1 medical social worker) and 4 students majoring in medicine and healthcare (e.g. therapy, nursing). Our interview protocol is composed of five main parts. The list of interview questions for a semi-structured interview can be found in Appendix. Table \ref{tab:interview_questions}

First, we asked participants to describe their work environment, how they build a trustworthy relationship with their colleagues, and how they discuss uncertain cases. We included this first question to understand any practices or aspects that should be considered for them to have trustworthy interaction with an AI system (Appendix. Table \ref{tab:interview_questions}). \pinkbox{P13} and \pinkbox{P14} shared the responses based on their experience of working as a therapy assistant in a hospital. \pinkbox{P15} elaborated on the experience of working with medical students and \pinkbox{P16} described the experience of working as a part-time nurse.

Next, we introduced the context and primary application of this study (i.e. physical stroke rehabilitation and a decision support system) (Figure \ref{fig:tutorial_context1},  \ref{fig:tutorial_apps}, and \ref{fig:tutorial_context2}) and explained inputs and an output of an AI-based decision support system for rehabilitation assessment and how it can be developed and operated on a new case using onboarding tutorial materials (Figure \ref{fig:tutorial_ai}). Also, a participant was asked to review inputs of a new case to an AI model and guess expected AI outputs.

Third, we described the dataset (Figure \ref{fig:tutorial_data}), evaluation metrics, and AI performance along with therapists' agreement levels (Figure \ref{fig:tutorial_performance}). The therapists' agreement levels indicate how well annotations of a secondary therapist are aligned with ground truth scores and could provide a reference on how well our AI model performs. After describing the dataset, evaluation metrics, and AI performance, we asked the participants without experience of using AI systems and AI explanations for their practices (1) if they have a particular baseline performance of an AI model that is required and (2) if they have any specific conditions or edge cases \cite{cai2019hello} that should be included in the dataset or which an AI model should be evaluated or good for considering using AI or for their trust usage (Appendix. Table \ref{tab:interview_questions}). Note that we did not ask the second question to participants (\yellowbox{P11 - P12} and \pinkbox{P13 - P16}), who do not have extensive experience in stroke rehabilitation as a therapist.

Fourth, after introducing the motivation of an AI explanation (Figure \ref{fig:tutorial_xai}) and three widely used AI explanations (i.e. feature importance (Figure \ref{fig:tutorial_feat}), counterfactual (Figure \ref{fig:tutorial_counter}), and prototype/example-based (Figure \ref{fig:tutorial_example}) explanations), we asked participants to rank which AI explanations are useful to support onboarding (i.e. when a user initially starts reviewing and understands AI performance) and decision support (i.e. when a user starts reviewing AI outputs for their decision making) (Appendix. Table \ref{tab:interview_questions}). Finally, we asked them to share any comments on how to improve onboarding with an AI system and validating an AI output during decision support (Appendix. Table \ref{tab:interview_questions}).

Throughout the semi-structured interviews, after reviewing onboarding tutorial materials, we asked the participants whether they followed the tutorial materials or had any clarification questions and asked them to provide any feedback on materials. All interviews were conducted remotely on a video conference platform and recorded for data analysis. Each interview lasted between 60 to 80 minutes. The participants were compensated for their participation based on the rate recommended by the domain experts. 

\subsection{Data Analysis}
We transcribed all 17.23 hours of interview recordings into text data for thematic analysis \cite{braun2012thematic}. We utilized both deductive and inductive thematic analysis approaches \cite{braun2012thematic,gale2013using} to analyze our interview data. We first selected codes based on our interview questions for our research problem: practices of rehabilitation and building a trustworthy relationship, comments on AI outputs, a development pipeline, the dataset, and the minimum performance, and comments on the AI explanations, onboarding with AI, and AI-assistive decision-making. Three researchers including one who facilitated interviews then independently coded the transcript data to generate 519 codes. We refined the initial codes while discussing disagreements and ambiguities in the codes \cite{mcdonald2019reliability,braun2012thematic,gale2013using} through iterative sessions. Following the practice of a reflective analysis that collaboratively shapes codes through discussion for consensus of codes \cite{braun2012thematic,mcdonald2019reliability}, we did not calculate inter-rater reliability.
After coding, we grouped similar codes to identify and conceptualize higher-level themes through affinity diagramming. Overall, this process yielded five high-level themes, twenty one second-level themes, and eighty five third-level themes (Appendix. Table \ref{tab:high-low-themes}). In Section \ref{sect:results}, we summarize our findings that broadly corresponds to the higher-level themes that we identified from our interview data.

\section{Results}\label{sect:results}

\subsection{Clinical Practices of Therapy \& Trustworthy Relationships}\label{sect:results_practices_trust}
\subsubsection{Clinical Practices}
Our participant therapists were from various settings: outpatient clinic (4); inpatient rehabilitation (3); skilled nursing facility (2); and home care (2). For a holistic understanding of a patient ({P5}, {P7}, {P14}), these settings typically have different sizes of interdisciplinary teams ranging from 2 to 24 team members (e.g. doctors, nurses, occupational therapists, physiotherapists, therapist assistants, and speech therapists). In most cases, they work in pairs or with a team of colleagues to \textit{``brainstorm how to assess patient’s status''} ({P9}) and \textit{``understand the needs of a patient to coordinate therapy sessions''} ({P8}). 

When handling a case, participants might get referrals of experienced colleagues ({P4}) for discussions. Otherwise, they determine which colleagues will be adequate by checking the following  aspects through their previous \textit{``interactions with them''} ({P8}): (i) \textbf{experiences and knowledge} (all participants), (ii) \textbf{quality of work} ({P10}), and various \textbf{soft skills} ({P5}, {P10}, {P12},{P14},{P15}). 
Specifically, they will approach more senior and experienced colleagues with hard skills, such as a corresponding specialty ({P1,P5,P7,P8,P10},{P14,P15,P16}) or experience with similar cases ({P3,P6}) for \textit{``receiving task-oriented advice''} ({P16}). In addition, another hard skill that they can check is whether colleagues provide client-centered quality work or not ({P10},{P11},{P13}). Lastly, they also highly consider soft skills of their colleagues, such as whether their colleagues are \textit{``more approachable''} ({P10}) and comfortable to share and discuss ({P14,P16}) to determine appropriate peers or colleagues for discussion. 

\begin{figure*}[ht]
\centering
  \includegraphics[width=0.5\columnwidth]{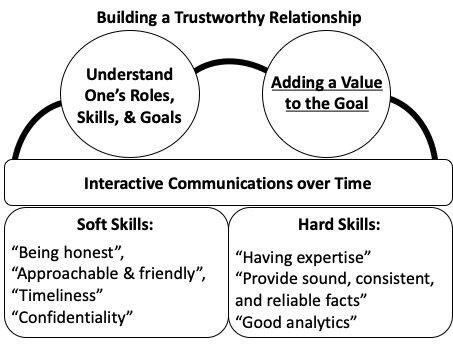}
\caption{Characteristics and skills of a trustworthy colleague and a process to build a trustworthy relationship with colleagues: a trustworthy relationship requires soft and hard skills to have interactive communications over time to understand colleagues' roles, skills, and goals and add a value on their goals.}\label{fig:trustworthy-relationship}
\end{figure*}

\subsubsection{Process to Build a Trustworthy Relationship}\label{results:build_relationship}
Participants commented that building a trustworthy relationship (Figure \ref{fig:trustworthy-relationship}) requires soft and hard skills to have \textbf{interactive communications over time}, in which one \textbf{understands colleagues' roles, skills, and goals}, but also checks whether a colleague \textbf{adds a value on common goals} (P7,P9,P11,P12).
For having a common goal, participants mentioned the importance of interacting with each other to know educational backgrounds and strategies (P4,P9) and \textit{``understand each other’s needs and roles''} (P12). As they \textit{``learn from each other''} (P3) and \textit{``show that they can help and do the work well''} (P5) over multiple interactions or sessions (P1,P2,P6), they build \textit{``a deeper trustworthy relationship with colleagues''} (P1). 

\subsubsection{Skills \& Characteristics of Trustworthy Colleagues}
Participants shared several skills and characteristics of trustworthy colleagues for interactive communication. First, soft skills, such as being honest, approachable \& friendly (P1,P7,P5,P10), and timely (P8,P15) are important to have iterative communications and identify common goals for building a trustworthy relationship. Given common goals, trustworthy colleagues also require hard skills to demonstrate one's expertise through \textit{``provid[ing] sound, consistent, reliable facts''} (P7). In addition, having good reputation with others (P5,P15) and keeping the confidentiality of information (P12) are important to build a relationship over time.

\subsection{Understanding and Information Needs of AI}\label{sect:results_info_ai}
Overall, participants understood the high-level ideas of an AI system for rehabilitation after presenting the tutorial materials. However, participants \textit{``are not clear about how the statistical methods work in detail''} (P12). 

When participants were asked to review inputs of an AI-based system and guess its outputs, participants had diverse ranges of correct guesses: among 16 participants, three participants (P1,P10,P13) correctly guessed all three outputs, nine participants (P3,P4,P5,P6,P7,P9,P11,P14,P15) correctly guessed two out of three outputs, two participants (P2,P12) correctly guessed one out of three outputs, and two participants (P8,P16) incorrectly guessed all outputs. When we analyzed the correlation between participants' normalized scores on technology experiences and their normalized scores on guessing AI outputs by computing Pearson’s correlation coefficient ($r$). We found that the coefficient value ($r$) of -0.31 and the p-value of 0.23, which indicates \textbf{a weak correlation between normalized scores on technology experiences and guessing AI outputs}.

While going through the tutorial materials, participants asked \textbf{questions about functional, operational, development, and evaluation aspects of AI}. 

\subsubsection{Functional \& Operational Aspects}
For functional aspects, participants (P1,P2,P4,P5,P6,P9) questioned how AI processes data and they could interact with AI. Specifically, participants wondered if AI can automatically identify incorrect data without therapists’ input (P1,P5). In addition, participants questioned how AI will operate if input data is slightly off from normal ones (P1,P3,P9,P10).
Participants also asked \textit{``whether AI can be evolved with new data''} (P7) and \textit{``inputs from therapists''} (P10) to have \textit{``an adaptive goal/normality''} (P10).

In addition, participants inquired about the easiness of setting a device/system (P1,P4), \textit{``how fast the system can process data to provide an assessment''} (P5),  and \textit{``how it might be repaired and maintained''}(P7). 

\subsubsection{Development \& Evaluation Aspects}
Participants questioned about how to interpret a confusion matrix (P11,P16), \textit{``any benchmark to determine whether AI is trustworthy or not''} (P7), \textit{``whether AI can become more accurate as it has a larger sample size''} (P15), and an evaluation setting (e.g. \textit{``whether AI is being piloted in a hospital''} - P14).

When it comes to the dataset, participants suggested additional factors that can be considered to expand the dataset. Overall, participants considered that the usage of the clinically validated functional assessment scores (e.g. fugl meyer assessment \cite{gladstone2002fugl}) is good to characterize and recruit post-stroke survivors for data collection. In addition to this functional assessment score, participants mentioned about distinguishing post-stroke survivors by other clinically relevant factors: the stage and severity of stroke (P2,P4,P5,P7,P10,P14), spasticity and muscle tone (P3,P7,P8), \textit{``the status of finer motor functions''} (P10), \textit{``cognitive status''} (P9) that a lot of post-stroke survivors struggle with. 

In addition, participants suggested expanding the dataset to make a more balanced distribution of sex (P2,P4,P6,P11,P12,P12), age (P2,P5,P8,P9,P11,P13), race (P4) as different sex, ages, and races might have \textit{``different characteristics and factors that affect their recovery''} (P2) and consider how to address \textit{``possible bias on labels''} (P1).

\subsection{Context-specific Required AI Performance \& Evaluations}
Most participants had \textbf{difficulty with enumerating how much percentage is sufficient} as they are not familiar with this metric (P4,P5,P10,P15) and \textit{``wondered if there is any industry standard to determine a good score''} (P3). \textit{``When AI has a lower performance, we[they] won’t likely to use it''} (P5). Thus, participants still described that it would be good to achieve high accuracy (P3,P15) ranging from 80\% (P2,P8,P12), 85\% (P3,P9), to 90\% (P6,P7,P10,P13) even if they \textit{``do not know whether it is even possible''} (P10). 

Participants mentioned the \textbf{importance of having context-specific required performance and a way to interpret the meaning of numbers}. P11 suggested a context-specific, desirable performance of AI: \textit{``when the AI is used as a reference, 70 - 80\% might be enough as therapists would make final assessment''} and \textit{``when the AI is being used independently, I expect a much higher score at least 95\%''}.  

Participants \textbf{recommended (i) presenting a benchmark performance and (ii) having contextualized and iterative evaluations}. \textit{``Even if we[they] are given this numerical performance value, we[they] have difficulty to interpret an actual meaning about what it implies''} (P10).  Participants considered presenting a benchmark performance score (e.g. therapists’ agreement) would be useful to \textit{``build an initial guide for comparison''} (P15) and check whether AI is reliable or not (P8,P14). Also, they suggested contextualizing evaluations by reporting how well AI will perform on the following aspects: common symptoms (P5,P6) (e.g. \textit{``assessing the range of motion''} - P1) and difficult tasks (P4,P7,P8,P9,P10,P13) (e.g. muscle tone, pelvis, scapular shoulder compensation, spasticity, finer finger motions, knee replacement, and gait patterns), \textit{``borderline cases''} (P3), and uncontrolled situations and data (P3,P5), such as presentation of other people than a patient and different setups of a system and a camera.

In addition, participants mentioned that \textit{``having trials to see how AI works would be necessary''} (P10) and \textit{``investigate how to improve it''} (P16) instead of presenting a number once as they \textit{``need time to build trust with AI as we build trust with our colleagues''} (P10).

\subsection{Understanding and Information Needs of AI Explanations}\label{sect:results_info_xai}
Overall, participants \textbf{desired descriptions on how they can leverage AI explanations}. Also, a few participants \textbf{inquired underlying processes of identifying AI explanations} and technical terms in visualizations. 

After asking a few clarification questions, participants understood the high-level ideas of three AI explanations (i.e. feature importance, counterfactual, prototype/example-based explanations). For a feature importance explanation, most participants followed the high-level concept in the first place. However, when some participants (P2,P3,P10) first encountered a local bar plot of SHAP values for each feature  \cite{lundberg2017unified} (Figure \ref{fig:tutorial_feat}), they asked how to read and interpret the graph (i.e. the meaning of blue and red plots, presented values). 

For a counterfactual explanation, some participants can follow the concept of a counterfactual explanation by correlating their similar practice (P2,P4), \textit{```familiarization', in which we determine why a client cannot get a correct position''} (P4). However, even if some participants  understood the high-level ideas of a counterfactual explanation, they were confused about how these explanations can be used to validate an AI output (P3,P10,P14,P15). 

For an example-based explanation, most participants considered that it is \textit{``easier to understand and review''} (P1). Some participants asked questions on the procedures of generating an example-based explanation: \textit{``how similar items are identified''} (P2) and \textit{``if the clothing color of a patient affects which similar cases will be identified''} (P2); \textit{``whether an AI model generates a similar sample''} (P9); how many samples are needed to define a prototype or find relevant cases (P2,P8). In addition, after reviewing the visualization of embedding spaces of samples (Figure \ref{fig:tutorial_example}), P3 asked about the meaning of the axes to project samples.

\subsection{Usefulness Ranking of AI Explanations for Onboarding and Decision Support}\label{sect:results_xai_rankings}
\begin{figure*}[ht]
\centering
  \includegraphics[width=1.0\columnwidth]{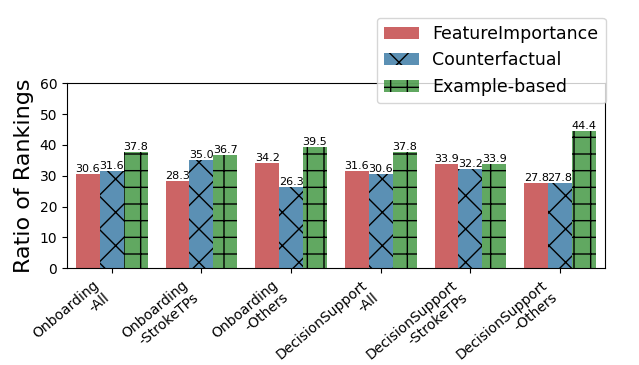}
\caption{Ratio of Rankings on the Perceived Usefulness of AI Explanations (Feature Importance, Counterfactuals, and Example-based) for Onboarding and Decision Support with AI from all participants, therapists with experience of stroke rehabilitation, and others (other health professionals and students majoring in medicine and health).}\label{fig:results-rankings}
\end{figure*}

Figure \ref{fig:results-rankings} summarizes the overall ratio of rankings on three AI explanations for onboarding and decision support using data from all participants, therapists with experience in stroke rehabilitation, and other participants.

\textbf{For onboarding, both therapists and other health professionals and students considered an example-based explanation as the most useful}. Therapists considered a counterfactual explanation (35.0\%) and a feature importance explanation (28.3\%) as the second and the third most useful. Other health professionals and students ranked a feature importance explanation (34.2\%) and a counterfactual explanation (26.3\%) as the second and the third most useful.

\textbf{For decision support, therapists considered both an example-based explanation and a feature importance explanation as equally useful (33.9\%) the most} and a counterfactual explanation (32.2\%) as the third most useful. \textbf{Other health professionals and students ranked an example-based explanation as the most useful} (44.4\%) and both a feature importance explanation and a counterfactual explanation as the second most useful ones (27.8\%).

When participants ranked the usefulness of AI explanations, they considered whether \textit{``it provides clinically relevant and useful information''} (P9) and whether it is \textit{``easy to understand and interact''} (P10). 

Among 16 participants, eight participants (P1,P4,P7,P9,P11,P12,P14,P15) had the same rankings on three AI explanations for onboarding and decision support. They considered that \textit{``both onboarding and decision support processes require the same process of reviewing and validating AI outputs''} (P7), in which they check how well \textit{``AI provides clinically relevant information''} (P9). Thus, they mentioned that \textit{``the usage of AI explanations would be similar''} (P11) for both onboarding and decision support. 

In contrast, eight participants (P2,P3,P5,P6,P8,P10,P13,P16) had different rankings on three AI explanations for onboarding and decision support. They differentiated the processes of validating AI outputs for onboarding and decision support phases. For instance, they described that they want to \textit{``validate how well AI outputs are useful to support a patient-specific assessment''} (P8) for decision support while they aim to \textit{``briefly validate the correctness of AI outputs to develop a trust with AI''} (P10) for onboarding.

\subsubsection{Feature Importance Explanation}
For a feature importance explanation, participants considered that it is useful to review only important features as \textit{``reviewing all features can be time-consuming''} (P7). However, participants also described the limitation of a feature importance explanation that it might identify features that are not important (P3,P4,P14) and less correlated with an outcome (P2,P16), which \textit{``might not be applicable and useful for the practice''} (P9). 

Some participants mentioned that \textit{``reviewing which features AI identifies as important''} (P6) is useful to understand how the AI works (P12,P13,P16) and \textit{``check the strength and limitations of AI''} (P15) for onboarding. In contrast, some participants elaborated that a feature importance explanation is more useful to validate an individual assessment for decision support than onboarding (P5,P6,P16). 

\subsubsection{Counterfactual Explanation}
For a counterfactual explanation, participants described that it is useful to review how to update patient’s movement features to flip an AI output (P7,P11), which is \textit{``what we want to achieve for our patient (e.g. how to make a patient with compensation to not have compensation)''} (P9). However, participants found it is difficult to understand (P1,P10) and time-consuming to get used to and validate whether counterfactual explanations make sense or not (P1,P5,P12,P16).

Some participants mentioned that reviewing counterfactual explanations \textit{``how features need to be changed to flip an AI output''} (P3) is helpful to understand how AI defines the medical conditions (e.g. compensation) (P6,P13) and understand the accuracy and performance of an AI model (P3,P8) for onboarding. Other participants described that counterfactual explanations \textit{``bring richer insights on both normal and abnormal conditions than other explanations that provide information on a single condition''} (P2) and are \textit{``useful to validate an AI output''} (P2) and \textit{``applicable in practice''} (P3) for decision support.

\subsubsection{Prototype/Example-based Explanation}
For an example-based explanation, participants valued that it is useful to review relevant samples to a client/patient’s condition instead of searching a whole dataset (P5,P13,P14) or \textit{``relying on our memory to remember all past cases''} (P9) and easy to understand to validate whether AI is correct or not (P1,P10,P12,P16). However, some participants mentioned that as individual post-stroke survivors are very different and specific, reviewing relevant examples might not be very useful (P7,P8,P16). They also wondered whether AI might have enough samples to provide relevant samples (P3,P8). 

Participants described the usefulness of an example-based explanation for onboarding as it shows how an AI model defines a clinical concept/symptom (P5,P6), and it is \textit{``easier to interpret than others''} (P6). In addition, reviewing a pool of samples is useful to draw relevant conclusions (P2,P3) and \textit{``confirm the validity of an AI output''} (P13) for decision support.

\subsection{Suggestions to Improve Onboarding and Decision-Making with AI}
For improving the understanding of the strengths and limitations of AI during onboarding and human-AI collaborative decision-making, participants recommended \textbf{(i) communicating benchmark information and the benefits of AI and (ii) a trial period to interact with AI for calibrating user trust, refining an objective of AI, and tuning AI with user feedback}.

\subsubsection{Communicating Benchmark Information \& Benefits of AI}
When we asked the participants to define a desired performance value of an AI model, they had difficulty with enumerating what it is desirable performance. Participants considered that \textit{``presenting benchmarkable information''} (P7), such as how much AI matches with therapists’ agreement (P2,P3,P5,P6,P14) or characterizing the performance on different medical conditions (P4,P6,P14) is necessary to better estimate a desirable performance for trustworthy AI (P3,P5). 

Participants also suggested that describing \textit{``the benefits of using AI''} (P4). As most participants had difficulty with contextualizing what a specific score of an evaluation metric means, they considered that it would be more effective to communicate  the benefits that they can easily understand. The example of these benefits of AI include \textit{``how much time can be saved''} (P5), \textit{``how well they can improve their decision than a therapist alone''} (P5), or whether it can support a better health outcome for patients (P2,P4). 

\subsubsection{Interaction Trials to Calibrate User Trust, Refine AI Objective, and Tune AI with Feedback}\label{sect:dis_refine}

Participants considered that demonstrating the strengths of AI through presenting a numerical value on an evaluation metric is still important. However, as a numerical value might not be sufficient to show the whole picture of AI performance, they also suggested providing a trial period with multiple interactions (P1,P5,P13,P14,P16) similar to how they build a trustworthy relationship with their colleagues over time (Section \ref{results:build_relationship}). By directly interacting with AI and observing how AI performs (P1,P12,P14,P16), participants considered that they can \textit{``check whether AI outputs are similar to what therapists consider''} (P7), understand the strengths and limitations of AI, and determine \textit{``whether AI is really helpful or not''} (P5). 

For a trustworthy relationship with colleagues, participants described the necessity of interactive communications over time to align a common goal (Table \ref{results:build_relationship}). Along this line, participants wondered \textit{``if AI can have an adaptive goal to set the notion of a correct movement based on patient's status''} (P10). In addition, participants desired a way to provide feedback to AI and update it accordingly (P5,P7) when any limitations of AI are identified and periodically refine AI with relevant, up-to-date data (P7,P9,P15) for more trustworthy interactions with AI.

\subsubsection{Other Considerations: Periodic Audits, Multi-sites Validations, and Easy Setups \& Usage}\label{sect:findings_considerations_audit}

In addition to communicating the competence and benefits of AI and interaction trials, participants described other important factors to consider using AI in practice and make AI more trustworthy. 
First, participants considered that it is necessary to have periodic, internal \& external audits (P2,P14,P16) \textit{``even after passing the onboarding or a trial period''} (P12). These audits refer to whether \textit{``AI can effectively provide relevant information''} (P7), \textit{``therapists correctly use the system''} (P12), and refining AI (Section \ref{sect:dis_refine}), but are not limited to these \cite{raji2020closing,mokander2021ethics,munoko2020ethical}.

 Participants also considered that having validations with multiple colleagues in multiple sites (P1,P12) would be helpful to consider using AI. As \textit{``each therapist has different educational backgrounds''} (P16) and \textit{``each hospital has different cultures''} (P14), participants wondered how an AI-based system can be deployed and applicable in various settings.

Finally, as \textit{``time-consuming logistics and setups would be a big barrier''} (P8) to consider using an AI-based system, these systems should be easy to set up and use as health professionals do not have much time in a clinical setting (P1,P5,P7,P10).

\section{Discussion}
In this section, we highlight key takeaways on the usefulness and value of tutorials on AI and AI explanations to improve communicating the strengths and limitations of AI for onboarding and human-AI collaborative decision-making. 
In addition, we discuss  design recommendations and future research for more effective onboarding with AI and AI explanations and human-AI collaborative decision-making: 1) improving tutorials on AI and AI explanations, 2) AI explanations for onboarding and interactive communications with AI, 3) measuring the level of understanding of AI and AI explanations, 4) beyond presenting a numerical, traditional performance metric, 5) goal alignment between users and AI and refining AI, and 6) audits to build a reputation of AI.

\subsection{Values \& Limitations of Tutorials on AI and AI Explanations}
An AI model card \cite{mitchell2019model} aims to provide a useful way to provide the essential facts of AI models in a structured way. However, our results showed that an AI model card is not sufficient to support onboard health professionals with AI. Even before presenting an AI model card, as most of our participants haven’t used much AI, they are clueless about AI (e.g. how it works and helps or what the limitations are) (P1,P5,P6). Without any tutorials, \textit{``it will be difficult for the first-time user without a technical background to use it''} (P4). Thus, they described the importance of having tutorials in simple languages (P6,P16) so that they can \textit{``make use of AI in a short time''} (P1). 

Much in the way that participants (i.e. health professionals and students majoring in medicine and health) desired the characteristics of being "friendly" and "approachable" as a trustworthy colleague (Figure \ref{fig:trustworthy-relationship}), our findings echo the needs to contextualize technical terminologies and make them "friendly" and "approachable" to improve the understanding of AI for AI-assisted decision-making \cite{shen2020designing,cai2019hello,kuo2023understanding}.

Our participants mentioned that they learned a lot about what’s behind AI and AI explanations (P8,P10) through our onboarding tutorial materials. They highlighted the importance of educating and \textit{``introduce[ing] about AI and AI explanations so that we[they] can have a better understanding on these to make the effective usage''} (P10). However, we found that some participants still asked clarification questions on AI or/and AI explanations after introducing our onboarding materials. In addition, even if participants developed understanding of AI and AI explanations, some participants were not clear how they can determine when AI is `ready' to be used.  \textit{``We[Users without technical backgrounds] do not necessarily think about inspecting whether AI is trustworthy or not''} (P2) and they \textit{``do not have any experiences to check the validity of AI''} (P1). Thus, it is necessary to educate and present a way (e.g. AI explanations) for them to effectively onboard with AI but also inspect AI outputs (e.g. \textit{``communicating why AI generates a certain output and how to validate it''} - P3). 

We expect participants might need additional interactive tutorials on the development pipeline of an AI model (e.g. how it processes data and metrics) and how to interact with and make use of AI and an AI explanation. Along this line, further research could explore how onboarding materials can be effectively delivered through interactivity, the choice of visual cues, and data visualizations \cite{crisan2022interactive}.

\subsection{Design Recommendations}
\subsubsection{AI Explanations for Onboarding and Interactive Communications with AI}
\hfill\\
AI explanations that aim to describe the behavior of the entire AI model or a specific AI output \cite{doshi2017towards,lakkaraju2020explaining,wang2019designing,abdul2018trends} have been increasingly explored to allow a user to understand when an AI model is right or wrong and can be trusted for AI-assisted decision making \cite{lai2021towards,wang2021explanations,bussone2015role}. However, these AI explanations have been mainly explored to provide insights on an AI output during the inference phase of an AI model for user's decision support \cite{poursabzi2021manipulating,cai2019human,zhang2020effect}. There have been limited explorations on how AI explanations can be used to support user's onboarding phase before moving to AI-assisted decision making phase. 

When we asked the participants to rank the usefulness of three AI explanations for onboarding and decision support phases, some participants differentiated the characteristics of onboarding and decision support tasks. Our results (Section \ref{sect:results_xai_rankings} and Figure \ref{fig:results-rankings}) show that the usefulness rankings of AI explanations are different depending on the tasks (e.g. onboarding vs decision support) and also participants had different strategies of using AI explanations on a task. Aligned with previous research that describes the necessity of characterizing tasks and stakeholders \cite{suresh2021beyond}, our study discusses a research problem of how AI explanations can be designed and used for onboarding phases. In addition, as interactive communications are critical to build a trustworthy relationship with colleagues Figure \ref{fig:trustworthy-relationship}), a further study is required to explore how AI explanations can serve as a tool/medium for health professionals to have interactive communications with AI.

\subsubsection{Measuring Understanding of AI and AI Explanations}
\hfill\\
Our study also uncovered challenges and needs in measuring the level of understanding of AI and AI explanations for more effective onboarding with AI/ML systems. When we recruited participants, we asked participants to respond with their experience of recent technologies (i.e. computer/laptop, activity tracker, virtual voice assistant, unmanned convenient store, and autonomous vehicle) based on the questions designed by the Center for Research and Education on Aging and Technology Enhancement (CREATE) \cite{czaja2006factors} that aim to measure and profile older adults' experience with technology \cite{beer2012domesticated}. However, one's experiences and higher exposures to recent technologies including AI applications do not necessarily mean that he or she will also have a good understanding of AI and AI explanations. For instance, P8 rated own technology experience as 6.2 out of 7 and P16 had 4.0 out of 7 but both P8 and P16 incorrectly guessed AI/ML outputs (Table \ref{tab:demographics}). In contrast, P10 and P13 had technology experience scores of 3.2 and 3.8 respectively, which is lower than those of P8 and P16. Both P10 and P13 correctly guessed all AI/ML outputs (Table \ref{tab:demographics}). Along this line, we found that normalized scores of technology experiences are not correlated with the normalized scores of guessing AI outputs (Section \ref{sect:results_info_ai}). 

In addition, even if our study explores the number of correctly guessing AI/ML outputs to measure participants' understanding of AI, this measurement is still limited. For instance, we observed that participants including ones who guessed all AL/ML outputs correctly asked clarification questions on AI and AI explanations. (Section \ref{sect:results_info_ai} and \ref{sect:results_info_xai}). 

Overall, our study results show that participants' technology experiences or their number of correctly guessing AI/ML outputs are not necessarily linearly correlated with their capabilities to understand how an AI operates nor how they appreciate the AI explanations. 
Thus, it is important to further explore how we can measure user's understanding of AI and AI explanations in a metric similar to how the previous studies assess the computer proficiency \cite{boot2015computer} and investigate what level of a metric indicates when a user has sufficient understanding to interact with AI and AI explanation.

\subsubsection{Beyond a Numerical Traditional Performance Metric}
\hfill\\
Our study results show the limitation of presenting the value of a traditional evaluation metric to either communicate the overall performance of an AI model or take a specific action (e.g. determining whether to deploy an AI-based system or not). Unlike participants' practice which they have interactive communications with colleagues over time to align a common goal and check the abilities of colleagues to add value to the common goal for their trustworthy relationships (Figure \ref{fig:trustworthy-relationship}), one-directional presentation of a numerical performance value does not provide the whole picture and understanding of AI for trustworthy onboarding and usage. 

For onboarding with an AI-based system, Cai et al. \cite{cai2019hello} recommended informing the overall performance of an AI-based system including its particular strengths and limitations. The Google's People + AI Guidebook \cite{GooglePAIR2019} recommends specifying a threshold value of a performance metric to take a specific action in the `User Needs + Defining Success' section. Given these recommended guidelines, we hypothesized that participants without technical background could discuss with AI/ML developers to review the reference performance of therapists' agreement level and specify a threshold value of a performance metric to determine whether an AI-based decision support system can be considered being used in the practice. 
However, even if we presented the performance of an AI model to assess three common performance components of rehabilitation assessment, our participants described the difficulty with understanding the overall performance of an AI model by reviewing a numerical value on a performance metric.  \textit{``If we are told that an AI model has 90\% accuracy, it might give a wrong mental model on AI performance as its performance might be changed in new cases''} (P11). Thus, participants desire a better way to \textit{``understand how well AI could perform on new data''} (P7), such as describing the benefits of using AI. However, the benefits of using AI cannot be communicated with a traditional evaluation metric. Thus, it is worthwhile to explore how to quantify the benefits of an AI-based system (e.g. clinical utilities) and describe its values to the end users in a more understandable way.  

\subsubsection{Goal Alignment between Users and AI and Refining AI}
\hfill\\
In addition, our study results suggest a gap between the objective of AI and the goal of a user. Although health professionals typically have a goal of improving patient's status, AI systems are trained to maximize the probability of replicating a therapist's assessment. For a trustworthy relationship with colleagues, participants described the necessity of interactive communications over time to align a common goal and add a value to it (Figure \ref{fig:trustworthy-relationship}). Along this line, participants wondered \textit{``if AI can have an adaptive goal to set the notion of a correct movement based on patient's status''} (P10). In addition, participants desired a way to provide feedback to AI and update it accordingly (P5,P7). When any limitations of AI are identified, they desire to periodically refine AI with relevant, up-to-date data (P7,P9,P15) for more trustworthy interactions with AI. For enabling trustworthy, human-AI collaborative decision-making, it would be critical to explore how to align an AI's goal with a user's goal and refine AI with the user's feedback \cite{lee2021human}.

\subsubsection{Audits to Build a Reputation of AI}
\hfill\\
As health professionals considered colleagues' reputation with others is an important factor in building a trustworthy relationship with their colleagues (Figure \ref{fig:trustworthy-relationship}), they also had a similar opinion that AI becomes more trustworthy if they hear more positive testimonials of colleagues from multiple sites (P4,P13,P14) (e.g. \textit{``colleagues can interpret AI outputs and reliably make use of them''} - P13). To this end, our study also highlights the values of audits on AI \cite{raji2020closing}. As mentioned in Section \ref{sect:findings_considerations_audit}, these audits can range from simply checking whether AI provides necessary information, refining AI with feedback \cite{lee2021human}. Also, the audit process can be monitoring and anticipating the potential negative impact of a system, designing mitigation or informing when to abandon the development and usage of an AI technology \cite{raji2020closing}. 
For onboarding with an AI-based system and integrating it in practice, our study also unveiled challenges and needs on how these AI-based systems for high-stake contexts can be audited and how we can support to educate people without technical backgrounds to participate in this process of deploying, onboarding, using, and auditing an AI-based system.

\subsection{Limitations}
Our study provides insights on information needs for onboarding with AI and AI explanations and discusses several existing gaps and areas to improve for more effective onboarding with AI and trustworthy human-AI collaborative clinical decision-making. However, our study is limited to introducing and presenting the onboarding tutorial materials to participants (i.e. health professionals and students majoring in medicine and health) by an online session. Our study does not observe how health professionals would initially interact with an AI system and AI explanations in a practical setting, which would bring richer and more useful insights on information needs to effectively onboard and interact with AI. 

In addition, our study has a limitation of generalizing the results as we mainly explore the research questions in the context of a single clinical decision-making tasks along with brief descriptions of an AI/ML model training and three AI explanations (i.e. feature importance, counterfactual, prototype/example-based). As the primary health professionals from the context of this study (i.e. rehabilitation therapy) are mostly females (e.g. around 62.7\% of therapists are female \cite{gender2023}), the participants of this study are mostly females (87.5\%). Also, our study does not involve a large number of participants even if such a small sample size is not unusual in similar previous works \cite{bussone2015role,cai2019hello}. A further in-situ study with other decision-making tasks and types of ML models and explanations is necessary for further generalizable insights on improving onboarding with AI and AI explanations. 

\section{Conclusion}
In this work, we contributed to an empirical study that explored how AI and AI explanations can be first introduced to health professionals and students majoring in medicine and health and identified information needs on AI and AI explanations for effective onboarding and trustworthy usage of AI.
Our study suggested the value of onboarding tutorial materials on AI and AI explanations and the necessity of 
designing AI explanations for improving onboarding and communications with AI. 
Also, our study highlighted the importance of exploring metrics to characterize the user's understanding of AI and AI explanations. 
In addition, our study discussed other considerations for effective onboarding and trustworthy human AI collaborative decision-making moving beyond describing a numerical traditional performance metric: presenting user-understandable benchmark information, interactive trials to communicate the practical benefits of AI, calibrate user trust, refine an objective of AI and AI with user feedback, and AI audits. Future research should explore how various types of AI/ML models and AI explanations on different contexts/tasks can be introduced to people without technical backgrounds.


\bibliographystyle{ACM-Reference-Format}
\bibliography{main}


\appendix

\section{Implementations of an AI model}\label{sect:appendix_aimodel}

We followed the previous research \cite{lee2019learning} to learn an AI model for rehabilitation assessment. Specifically, we processed the estimated joint positions of post-stroke survivors' exercises to extract various kinematic features. The kinematic features of the `Range of Motion' (ROM) include joint angles, such as elbow flexion, shoulder flexion, and elbow extension, and normalized relative trajectory (i.e. the Euclidean distance between two joints - head and wrist; head and elbow), and the normalized trajectory distance (i.e. the absolute distance between two joints - head and wrist, shoulder and wrist) in the x, y, and z coordinates \cite{lee2019learning}. The features of the `Compensation' include the normalized trajectories, which indicate the distances between joint positions of the head, spine, and shoulder in the x, y, and z coordinates from the initial to the current frame over the entire exercise motion \cite{lee2019learning}.

As previous research demonstrated the outperformance of a feed-forward Neural Network (NN) model to classify the quality of post-stroke survivors' motion \cite{lee2019learning}, we utilized the extracted kinematic features and labels of post-stroke survivors' exercises to implement a feed-forward NN model using Pytorch libraries \cite{paszke2019pytorch}. For the labels, we utilized the labels by the expert therapist, who conducted the clinically validated assessment test. We grid-searched various architectures (i.e. one to three layers with 32, 64, 128, 256, and 512 hidden units) and different learning rates (i.e. 0.0001, 0.0005, 0.0001, 0.005, 0.001) while training a feed-forward NN model with cross-entropy loss and the mini-batch size of 1 and epoch of 4. For training and evaluating the model, we utilized the leave-one-subject-out cross-validation, where we trained the model with data from all post-stroke survivors except one post-stroke survivor and tested the model with data from the held-out post-stroke survivor. The final model architectures and learning rates are three layers with 256 hidden units and 0.005 of the learning rate for the ROM, one layer with 16 hidden units and 0.0001 of the learning rate for the Smoothness, and three layers with 64 hidden units and 0.005 of the learning rate for the Compensation. The models achieved 82\% F1-score, 79\% F1-score and 77\% F1-score to replicate therapists' assessment on `ROM', `Smoothness', and `Compensation' components respectively.

\section{Details of the Study: Participants, Interviews, and Data Analysis}\label{sect:appendix_study}

\begin{table}[htp]
\centering
\caption{Detailed Demographics of Participants: Therapists who have experience in stroke rehabilitation (\bluebox{P1 - P10}) and other health professionals (\yellowbox{P11 - P12}) and students majoring in medicine or health (e.g. therapy, nursing) (\pinkbox{P13 - P16}).}
\label{tab:demographics_detailed}
\resizebox{\textwidth}{!}{%
\begin{tabular}{llllllll} \toprule
\textbf{PID} &
  \textbf{Sex} &
  \textbf{Age} &
  \textbf{Occupation} &
  \textbf{Setting} &
  \textbf{\# of yrs} &
  \textbf{Q. Tech Experience} &
  \textbf{Q. ML Outputs} \\ \midrule
\bluebox{P1} &
  Female &
  25 - 34 years &
  PhysioTherapist (PT) &
  Outpatient Clinic &
  7 &
  \begin{tabular}[c]{@{}l@{}}
  4.8 out of 7\end{tabular} &
  3 out of 3 \\ \midrule
\bluebox{P2} &
  Male &
  25 - 34 years &
  PhysioTherapist (PT) &
  Inpatient Rehabilitation &
  2 &
  \begin{tabular}[c]{@{}l@{}}
  5.2 out of 7\end{tabular} &
  1 out of 3 \\ \midrule
\bluebox{P3} &
  Male &
  25 - 34 years &
  PhysioTherapist (PT) &
  Home Care &
  8 &
  \begin{tabular}[c]{@{}l@{}}
  4.4 out of 7\end{tabular} &
  2 out of 3 \\ \midrule
\bluebox{P4} &
  Female &
  35 - 44 years &
  PhysioTherapist (PT) &
  Outpatient Clinic &
  11 &
  \begin{tabular}[c]{@{}l@{}}
  5.8 out of 7\end{tabular} &
  2 out of 3 \\ \midrule
\bluebox{P5} &
  Female &
  25 - 34 years &
  PhysioTherapist (PT) &
  Inpatient Rehabilitation &
  9 &
  \begin{tabular}[c]{@{}l@{}}
  5.4 out of 7\end{tabular} &
  2 out of 3 \\ \midrule
\bluebox{P6} &
  Female &
  45 - 54 years &
  PhysioTherapist (PT) &
  Skilled Nursing Facility &
  30 &
  \begin{tabular}[c]{@{}l@{}}
  5.8 out of 7\end{tabular} &
  2 out of 3 \\ \midrule
\bluebox{P7} &
  Female &
  35 - 44 years &
  Occupational Therapist (OT) &
  Outpatient Clinic &
  14 &
  \begin{tabular}[c]{@{}l@{}}
  5.4 out of 7\end{tabular} &
  2 out of 3 \\ \midrule
\bluebox{P8} &
  Female &
  35 - 44 years &
  Occupational Therapist (OT) &
  Homecare &
  11 &
  \begin{tabular}[c]{@{}l@{}}
  6.2 out of 7\end{tabular} &
  3 out of 3 \\ \midrule
\bluebox{P9} &
  Female &
  25 - 34 years &
  Occupational Therapist (OT) &
  Skilled Nursing Facility &
  6 &
  \begin{tabular}[c]{@{}l@{}}
  4.4 out of 7\end{tabular} &
  2 out of 3 \\ \midrule
\bluebox{P10} &
  Female &
  25 - 34 years &
  Occupational Therapist (OT) &
  Inpatient Rehabilitation &
  5 &
  \begin{tabular}[c]{@{}l@{}}
  3.2 out of 7\end{tabular} &
  3 out of 3 \\ \midrule
\yellowbox{P11} &
  Female &
  25 - 34 years &
  Speech Therapist &
  {Community outpatient} &
  5 &
  \begin{tabular}[c]{@{}l@{}}
  4.8 out of 7\end{tabular} &
  2 out of 3 \\ \midrule
\yellowbox{P12} &
  Female &
  25 - 34 years &
  Medical Social Worker &
  n/a &
  5 &
  \begin{tabular}[c]{@{}l@{}}
  4.0 out of 7\end{tabular} &
  1 out of 3 \\ \midrule
\pinkbox{P13} &
  Female &
  25 - 34 years &
  Student in Occupational Therapy &
  n/a &
  n/a &
  \begin{tabular}[c]{@{}l@{}}
  3.8 out of 7\end{tabular} &
  3 out of 3 \\ \midrule
\pinkbox{P14} &
  Female &
  25 - 34 years &
  Student in Speech Therapy &
  n/a &
  n/a &
  \begin{tabular}[c]{@{}l@{}}
  3.8 out of 7\end{tabular} &
  2 out of 3 \\ \midrule
\pinkbox{P15} &
  Female &
  18 - 24 years &
  Student in Medicine &
  n/a &
  n/a &
  \begin{tabular}[c]{@{}l@{}}
  3.8 out of 7\end{tabular} &
  2 out of 3 \\ \midrule
\pinkbox{P16} &
  Female &
  18 - 24 years &
  Student in Nursing &
  n/a &
  n/a &
  \begin{tabular}[c]{@{}l@{}}
  4.0 out of 7\end{tabular} &
  0 out of 3 \\ \bottomrule
\end{tabular}%
}
\end{table}

\begin{table}[htp]
\caption{List of questions for a semi-strucutured interview}
\label{tab:interview_questions}
\resizebox{\textwidth}{!}{%
\begin{tabular}{ll} \toprule
\multicolumn{1}{c}{\textbf{Parts of a Semi-Structured Interview}} &
  \multicolumn{1}{c}{\textbf{Prompt Questions}} \\ \midrule
\begin{tabular}[c]{@{}l@{}}Rehabilitation practices \\ \& trustworthy relationships\end{tabular} &
  \begin{tabular}[c]{@{}l@{}}How many colleagues you have \\ \& how frequently you interact with them for rehabilitation assessment?\end{tabular} \\
 &
  How do you build a trustworthy relationship with your colleagues? \\
 &
  When you have an uncertain case, how do you know a particular colleague will be good for discussion? \\
 &
  What aspects of your colleagues make them trustworthy? \\ \midrule
Intro to AI &
  \begin{tabular}[c]{@{}l@{}}Any questions/information needs about how the system operates\\ and the development and evaluation pipeline of an AI system\end{tabular} \\
 &
  \begin{tabular}[c]{@{}l@{}}We tried to collect data from post-stroke survivors with diverse FMA scores. \\ Do you have any particular post-stroke survivors that should be included in the dataset?\end{tabular} \\
 &
  \begin{tabular}[c]{@{}l@{}}AI system cannot be perfect. \\ Do you have a particular performance that is required for your trustful usage?\\ At least XX F1-score an AI model needs to achieve to consider using AI and your trustful usage if being used?\end{tabular} \\
 &
  \begin{tabular}[c]{@{}l@{}}Do you have any specific conditions (e.g. Full ROM, shoulder/trunk compensation) or edge cases \\ that AI should be evaluated and good at for consider using AI and your trustful usage if being used?\end{tabular} \\ \midrule
AI Explanations &
  \begin{tabular}[c]{@{}l@{}}Any questions or information needs about AI explanations \\ (e.g. Feature Importance, Counterfactual, Example-based)?\end{tabular} \\
 &
  \begin{tabular}[c]{@{}l@{}}Rank which AI explanations are the most useful to onboard with AI \\ and understand overall performance \& strengths/limitations of AI?\end{tabular} \\
 &
  \begin{tabular}[c]{@{}l@{}}Why do you consider that a particular explanation is useful or not to onboard with AI \\ and understand overall performance and strengths/limitations of AI?\end{tabular} \\
 &
  \begin{tabular}[c]{@{}l@{}}Rank which AI explanations are the most useful to make a decision with AI \\ and determine whether to trust an AI outcome or not?\end{tabular} \\
 &
  \begin{tabular}[c]{@{}l@{}}Why do you consider that a particular explanation is useful or not to \\ make a decision with AI and determine whether to trust an AI prediction or not?\end{tabular} \\ \midrule
Wrap-up &
  \begin{tabular}[c]{@{}l@{}}Any comments/suggestions on how we can address the following issues:\\ \textgreater how to determine whether AI has sufficient performance\\ \textgreater how to improve onboarding with AI and determine strengths/limitations of AI\\ \textgreater how to improve decision-making with AI and inspect whether AI outputs are trustful or not\end{tabular} \\ \bottomrule
\end{tabular}%
}
\end{table}

\begin{table}[htp]
\caption{The five high-level themes, twenty one second-level themes, and eighty five third-level themes of qualitative data analysis}
\label{tab:high-low-themes}
\resizebox{\textwidth}{!}{%
\begin{tabular}{lll} \toprule
\multicolumn{1}{c}{\textbf{High-level Themes}} &
  \multicolumn{1}{c}{\textbf{Second-level Themes}} &
  \multicolumn{1}{c}{\textbf{Third-level Themes}} \\ \midrule
Practices of Rehabilitation &
  Practices on Therapy &
  \begin{tabular}[c]{@{}l@{}}Teams, Assessment \& Therapy,\\ Meetings, Communications\end{tabular} \\
 &
  Practices on Uncertain Decision-Making &
  \begin{tabular}[c]{@{}l@{}}Meetings, Other Info, \\ Factors to identify colleagues (Referral, Expertises, Soft Skills)\end{tabular} \\
 &
  Process to Build a Trustworthy Relationship &
  \begin{tabular}[c]{@{}l@{}}Period of time, Understand role \& needs, \\ Share common goals, Add values, \\ Honest, Friendly, Listen \& Interact\end{tabular} \\
 &
  Characteristics of Trustworthy Colleagues &
  \begin{tabular}[c]{@{}l@{}}Personalities, \\ Knowledge \& Background, Consistent \& Reliable, \\ Work Ethics \& Timeliness, Confidentiality\end{tabular} \\ \midrule
Intro to AI &
  Guesses on AI outputs & -
   \\
 &
  Comments on the dataset of AI &
  \begin{tabular}[c]{@{}l@{}}Recruitment criterion\\ - Demographics of participants (age, sex, race, etc.)\\ - Stroke conditions (impact, spasticity, tone, finer motions, etc.)\\ Number of samples\\ Labels\end{tabular} \\
 &
  Required performance of AI &
  \begin{tabular}[c]{@{}l@{}}Context-specific thresholds, specific numbers, \\ Unclear meaning of numbers, \\ Reference\end{tabular} \\
 &
  Evaluation and edge cases of AI &
  \begin{tabular}[c]{@{}l@{}}Common symptoms, Difficult tasks, \\ Borderline cases,\\ Realistic, uncontrolled settings\end{tabular} \\
 &
  Questions about AI &
  \begin{tabular}[c]{@{}l@{}}Definitions, Operation speed, \\ Meaning of a confusion matrix and a confidence score, \\ How to perform and fail on a certain case\end{tabular} \\
 &
  Strengths of AI &
  Process data quickly, Reduce workload, Objective data \\
 &
  Limitations of AI &
  \begin{tabular}[c]{@{}l@{}}Troublesome and time-consuming, \\ Requires a specific setup, \\ Can't show me exact symptoms\end{tabular} \\
 &
  Suggestions on AI &
  \begin{tabular}[c]{@{}l@{}}Not familiar with how to onboard \& make a decision with AI\\ Need to be usable\\ Periodic audits,\\ Studies from multiple sites\\ Other rehabilitation-specific features\\ Others\end{tabular} \\ \midrule
AI Explanations &
  Comments/questions on AI explanations &
  Clarifications on the concept \\
 &
  \begin{tabular}[c]{@{}l@{}}Participants with the same/different rankings \\ on onboarding \& decision-support\end{tabular} &
   \\
 &
  Ranking of AI explanations for onboarding &
   \\
 &
  Ranking of AI explanations for decision-support &
   \\
 &
  Example-based explanations &
  \begin{tabular}[c]{@{}l@{}}Questions, Comments (General, Pros, Cons), \\ Onboarding, Decision-support\end{tabular} \\
 &
  Feature importance explanations &
  \begin{tabular}[c]{@{}l@{}}Questions, Comments (General, Pros,   Cons), \\ Onboarding, Decision-support\end{tabular} \\
 &
  Counterfactual explanations &
  \begin{tabular}[c]{@{}l@{}}Questions, Comments (General, Pros,   Cons), \\ Onboarding, Decision-support\end{tabular} \\
Onboarding with AI &
  Suggestions on onboarding &
  \begin{tabular}[c]{@{}l@{}}Values of tutorials, Beyond numbers,\\ Benchmark references, Demonstrate benefits,\\ Interactions, Referrals and testimonials,\\ Other factors\end{tabular} \\
AI-assisted decision-support &
  Suggestions on decision-support &
  \begin{tabular}[c]{@{}l@{}}Trial periods, \\ Update AI with feedback, \\ Other factors (e.g. setups, user-friendly)\end{tabular} \\ \bottomrule
\end{tabular}%
}
\end{table}

\end{document}